\documentclass[preprints,article,accept,moreauthors,pdftex]{Definitions/mdpi} 

\firstpage{1} 
\pubvolume{xx}
\issuenum{1}
\articlenumber{5}
\pubyear{2019}
\copyrightyear{2019}
\history{Received: date; Accepted: date; Published: date}

\usepackage{xcolor, bm}
\usepackage{subcaption}
\usepackage{amssymb, amsmath, bm, mathtools,textcomp,gensymb}
\usepackage{commath}
\usepackage[capitalise]{cleveref}
\usepackage[normalem]{ulem}

\newcommand{\eps}{\bm{\varepsilon}}

\newcommand{\PRM}{\boldsymbol{\vartheta}}
\newcommand{\CFG}{\boldsymbol{\varphi}}
\newcommand{\SNS}{\bm{s}} 		
\newcommand{\Signal}{\bm{F}} 	
\newcommand{\MS}{\bm{y}} 		

\newcommand{\CorLen}{\ell} 		

\newcommand{\COMMA}{\, ,}
\newcommand{\PERIOD}{\, .}

\newcommand{\PSPACE}{\mathcal{T}}
\newcommand{\YSPACE}{\mathcal{Y}}

\DeclareMathOperator*{\argmax}{argmax}

\newcommand{\EXPa}{size of the leading school}
\newcommand{\EXPb}{relative position}

\graphicspath{{./figures/}}

\Title{Optimal sensing for fish school identification}

\Author{Pascal Weber $^{1}$, Georgios Arampatzis $^{1}$,Guido Novati $^{1}$, Siddhartha Verma $^{2}$,  Costas Papadimitriou $^{3}$ and Petros Koumoutsakos $^{1,}$*}

\AuthorNames{Pascal Weber, Guido Novati, Georgios Arampatzis, Siddhartha Verma, Costas Papadimitriou and Petros Koumoutsakos}

\address{%
$^{1}$ \quad Computational Science and Engineering Laboratory, Clausiusstrasse 33, ETH Z{\"u}rich, CH-8092, Switzerland.\\
$^{2}$ \quad Department of Ocean and Mechanical Engineering, Florida Atlantic University, Boca Raton, FL 33431, USA and  Harbor Branch Oceanographic Institute, Florida Atlantic University, Fort Pierce, FL 34946, USA\\
$^{3}$\quad Department of Mechanical Engineering, University of Thessaly, Pedion Areos, GR-38334 Volos, Greece
}

\corres{Correspondence: petros@ethz.ch}

\abstract{
Fish schooling implies an awareness of the swimmers for their companions. In flow mediated environments, in addition to visual cues, pressure and shear sensors on the fish body are critical for providing quantitative information that assists the quantification of proximity to other swimmers. Here we examine the distribution of sensors on the  surface of an artificial swimmer so that it can optimally identify a leading group of swimmers. We employ Bayesian experimental design coupled with two-dimensional Navier Stokes equations for multiple self-propelled swimmers. The follower tracks the school using information from its own surface pressure and shear stress. We demonstrate that the optimal sensor distribution of the follower is qualitatively similar to the distribution of neuromasts on fish. Our results show  that it is possible  to identify accurately the center of mass and even the number of the leading swimmers using surface only information.
}


\begin{document}

\section{Introduction}
\label{sec:intro}

Fish navigate in their  environments by processing and adapting to cues from the surrounding flow fields. The flow environment is replete with mechanical disturbances
(pressure, vorticity)  that can convey information about the sources that generated them.  Fish swimming in groups have been found to process such external information and balance it with social cues \cite{Ward2008,Puckett2018}.
Fish may perceive mechanical disturbances in terms of surface pressure and shear stresses through neuromasts.
Early studies and experiments showed that the functioning of the lateral line is crucial for several tasks~\cite{Dykgraaf1933, Dijkgraaf1963}. 
Experiments with trouts in the vicinity of objects showed its importance for K\'arm\'an gaiting and bow wake swimming as well as energy efficient station keeping \cite{Bleckmann2012,Sutterlin2011}
Additionally, it was shown that the flow information allows to determine the cylinder diameter and flow velocity as well as the position relative to the generated K\'arm\'an vortex street by taking measurements of the flow~\cite{Akanyeti2011,Chambers2014}.
Using blind cave fish a study showed the importance of the lateral line to detect location and shape of surrounding objects and avoid obstacles \cite{vonCampenhausen1981,Hassan1989,Windsor2010a,Windsor2010b}. 
In another study the feeding behavior of blinded mottled sculpin was tested and it was found that they use their lateral line system to detect prey~\cite{Hoekstra1985}.
Also it was found that blind fish manage to keep their position in schools and lose this ability with disabled lateral line organ~\cite{Pitcher1976}. 
Further importance of the lateral line was shown for communicational behavior~\cite{Satou1994}, selection of habitats~\cite{Huijbers2012} and rheotaxis~\cite{Montgomery1997}. 

The lateral line consists of two functional classes of receptors: mechanoreceptors and electroreceptors. In this work we mimic the mechanosensory receptors, more specifically the sub-surface `canal' neuromasts and superficial neuromasts \cite{Coombs1988,Coombs1989}. The neuromast on the fish skin are used to detect shear stresses, where the ones residing in the lateral line canals are used to detect pressure gradients \cite{Denton1989, coombs2003, coombs2005, Bleckmann2008, Jiang2019}. Due to the filtering nature of the canals, the detection of small hydrodynamic stimuli against background noise is improved for the subsurface neuromasts~\cite{Engelmann2000}. 

The effectiveness and versatility of the lateral line organ inspired several bio-inspired artificial flow sensors~\cite{Kottapalli2012, Tao2012,  Asadnia2015, Kottapalli2018}. Arranging these sensors in arrays on artificial swimmers has attracted attention to transform underwater sensing~\cite{Yang2006, Yang2010, Kruusmaa2014, deVries2015, Strokina2016, Triantafyllou2016, Xu2017, Sengupta2019, Zhang2019}. Here, leveraging the intelligent distributed sensing inspired by the lateral line showed to be effective in robots moving in aquatic environments~\cite{Jezov2012, Yen2018, Krieg2019}. 

In order to better use and understand the capabilities of the artificial sensors several studies regarding the information content in the flow and optimal harvesting of this information were performed: The prevalence of information on the position of a vibrating source was shown to be linearly coded in the pressure gradients measured by the subsurface neuromasts~\cite{Curvcic2006}. 
Furthermore, it was shown that the variance of the pressure gradient correlated with the presence of lateral line canals~\cite{Ristoph2015}. A study with fish robots equipped with distributed pressure sensors for flow sensing and Bayesian filtering for estimating the flow speed, angle of attack, and foil camber~\cite{Zhang2015}. 
Other studies focused on dipole sources to develop methods to extract information and optimize the parameters of the sensing devices~\cite{Ahrari2015, Ahrari2017}. 
In a recent study artificial neural networks were employed to classify wakes from airfoils~\cite{Colvert2018}. In order to find effective sensor positions weight analysis algorithms were employed~\cite{Xu2019}. 

In this study, following earlier work for detection of flow disturbances by single obstacles,~\cite{Verma2019} we examine the optimality of the spatial distribution of sensors in a self-propelled swimmer who infers the number in a group of neighboring (here leading) fish and their relative positions. 
We combine numerical simulations of the two-dimensional Navier-Stokes equation and Bayesian optimal sensor placement to examine the extraction of flow information by pressure gradients and optimal positioning of sensors. 
The paper is organised as follows:
In \cref{sec:numMeth} we describe the numerical simulations and in \cref{sec:optimalSensor} the process of Bayesian optimal experimental design. We present our results in \cref{sec:results} and conclude in \cref{sec:conclusion}.

\begin{figure}
    \centering
	\includegraphics[width=0.9\columnwidth]{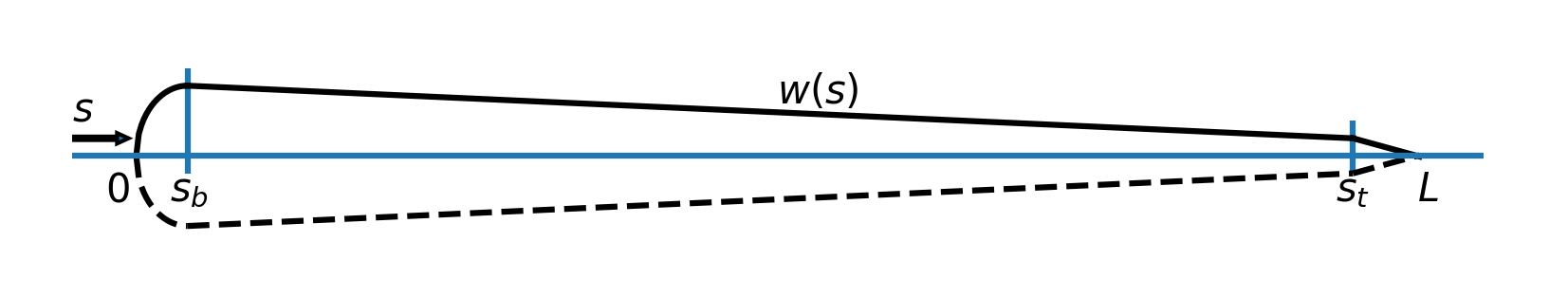}
	\caption{
	Parametrization of the fish surface as described in \cref{eq:fish:width}.
	}
	\label{fig:parametrisation}
\end{figure}

\section{Materials and Methods}

\subsection{Flow simulations}
\label{sec:numMeth}

The swimmers are modeled by  slender deforming bodies of length $L$ which are characterized by their half-width $w(s)$ along the midline \cite{Kern2006,Gazzola2011} 

\begin{equation}\label{eq:fish:width}
  w(s)=\begin{cases}
  \sqrt{2w_hs-s^2}, & 0
  \le s< s_b  \\
  w_h-(w_h-w_t)\left(\frac{s-s_b}{s_t-s_b}\right),& s_b\le s<s_t  \\
  w_t\frac{L-s}{L-s_t},& s_t\le s\le L  \PERIOD
  \end{cases}
\end{equation}
A sketch of the parametrization is presented in \cref{fig:parametrisation}.
Following \cite{kern2008anguilliform}, we use $w_h=s_b=0.04L$, $s_t=0.95L$ and $w_t=0.01L$. The swimmers propel themselves by performing sinusoidal undulations of their midline. This motion is described by a prescribed time-dependent parameterization of the curvature,

\begin{equation}
k(s,t)=A(s)\sin\left(\frac{2\pi t}{T_p}-\frac{2\pi s}{L}\right)\PERIOD
\end{equation}
Here $T_p = 1$ is the tail-beat period and $A$ is the undulation amplitude which linearly increases from $A(0) = 0.82/L$ to $A(L) = 5.7/L$ to replicate the anguilliform swimming motion described by~\cite{Carling:1998}.
Given the curvature along $s$ and a center of mass, the coordinates $\mathbf{r}(s,t)$ of the swimmer's midline can be computed by integrating the Frenet–Serret formulas~\cite{kern2008anguilliform}. In turn, the half-width $w(s)$ and the coordinates $\mathbf{r}(s,t)$ characterize the swimmer's surface.

The flow environment is described  by numerical simulations of the two-dimensional incompressible Navier-Stokes equations (NSE) in velocity-pressure ($\mathbf{u}-p$) formulation. The NSE are discretized with second order finite differences and integrated in time with explicit Euler time stepping.
The fluid-structure interaction is approximated with Brinkman penalization~\cite{Angot:1999, coquerelle2008vortex, Gazzola:2011} by extending the fluid velocity $\mathbf{u}$ inside the swimmers' bodies and by including in the NSE a penalization term to enforce no-slip and no-through boundary conditions,

\begin{align}\label{eq:NS}
\frac{\mathbf{u}^{k+1}-\mathbf{u}^{k}}{\delta t} = -\nabla p^{k} -(\mathbf{u}^{k}\cdot \nabla)\mathbf{u}^{k} + \nu \Delta \mathbf{u}^{k} + \sum_{i=1}^{N_s} \lambda\chi_i(\mathbf{u}^{k}_{s,i}-\mathbf{u}^{k}) \PERIOD
\end{align}
Here, $\nu$ is the kinematic viscosity, $\lambda = 1/\delta t$ is the penalization coefficient, $N_s$ is the number of swimmers, $\mathbf{u}^{k}_{s,i}$ is the velocity field imposed by swimmer $i$ (composed of translational, rotational and undulatory motions), and $\chi_i$ is its characteristic function which takes value 1 inside the body of swimmer $i$ and value 0 outside. The characteristic function $\chi_i$ is computed, given the distance of each grid-point from the surface of swimmer $i$, by a second-order accurate finite difference approximation of a Heaviside function~\citep{Towers2009}.
The pressure field is computed by pressure-projection~\citep{Chorin1968, Gazzola:2011},

\begin{equation}\label{eq:PresProj}
 \Delta p = \frac{1}{\delta t} \nabla \cdot \tilde{\mathbf{u}}^{k} - \frac{1}{\delta t} \sum_{i=1}^{N_s} \chi_i \nabla \cdot \tilde{\mathbf{u}}^{k}  \COMMA
\end{equation}
where $\tilde{\mathbf{u}}^{k} = \mathbf{u}^k -(\mathbf{u}^k \cdot \nabla) \mathbf{u}^k + \nu \Delta \mathbf{u}^k$. The terms inside the summation in \cref{eq:PresProj} are due to the non-divergence free deformation of the swimmers.

\subsubsection{Schooling formation}\label{sec:formation}

The tail-beating motion that propels forwards a single swimmer generates in its wake a sequence of vortices.  The momentum contained in the flow field induces forces to the swimmers that  fish in schooling formation must overcome to maintain their positions in the group ~\citep{novati2016synchronised}. In this study, we maintain the schooling formation for multiple swimmers by employing closed-loop parametric controllers. We increase or decrease the tail-beating frequency $T_{p,i}$ of each swimmer $i$ if it lags behind or surpasses a desired position in the direction of the school's motion,

\begin{equation}
   T_{p,i} = T_p (1 - \Delta x_i ) \PERIOD
\end{equation}
In order to straighten the school's trajectory we impose additional curvature $k_{C,i}$ along each swimmer's midline in order to minimize its lateral  deviation $\Delta y_i$ and its angular deflection $\Delta \theta_i$,

\begin{equation}\label{eq:PIDY}
k_{C,i} = [\Delta y_i, \langle\Delta \theta_i\rangle]_- + [\langle\Delta y_i\rangle, \Delta \theta_i]_- + [\langle\Delta y_i\rangle, \langle\Delta \theta_i\rangle]_- \PERIOD
\end{equation}
Here, $\langle\cdot\rangle$ defines an exponential moving average with weight $\delta t / T_p$, which approximates the integral term found in PI controllers and,

\begin{equation} [a, b]_- = \begin{cases}
\|a\|~b \qquad &\text{if} ~ab<0,\\
0 \qquad &\text{otherwise} \PERIOD
\end{cases} \end{equation}
The formulation in  \cref{eq:PIDY} indicates that if both the lateral displacement and the angular deviation are positive (or both negative) the swimmer will gradually revert to its position in the formation. Conversely, if $\Delta y_i$ and $\Delta \theta_i$ have different signs the displacement has to be corrected by adding (or subtracting) curvature to the swimmer's midline. 

\subsubsection{Flow sensors}\label{sec:sensors}

We distinguish two types of sensors on the swimmer body. The superficial neuromasts detect flow stress and the subcanal neuromasts pressure gradients~\cite{Kroese1992,Bleckmann2009,Asadnia2015}. From the numerical solution of the 2D Navier-Stokes equation we obtain the flow velocity $\mathbf{u}=(u,v)$ and the pressure $p$ at every point of the computational grid. The surface values of these quantities are obtained through a bi-linear interpolation from the nearest grid points.
We perform offline analysis by recording the interpolated pressure $p$ and flow velocity $\mathbf{u}$ in the vicinity of the body. We remark that we have neglected points near the end of the body to reduce the influence of large flow gradients that are generated by the motion and sharp geometry of the tail. The shear stresses are computed on the body surface using the local tangential velocity in the two nearest grid points. Moreover we compute pressure gradients along the surface by first smoothing thes pressure along the surface ${a}$ using splines implemented in  \textsc{SciPy} ~\cite{SciPy,Dierckx1975}.

\subsection{Optimal sensor placement based on Information gain}
\label{sec:optimalSensor}
In the present work a swimmer is equipped with sensors that are used to identify the size and location of a nearby school.
The  optimal sensor locations are identified using  Bayesian experimental design \cite{Huan2013} so that 
the information obtained from the collected measurements is maximized. 
We define the information gain as the  distance between the prior belief on the quantities of interest and the posterior belief after obtaining the measurements. Here, we choose as measure of the distance the Kullback-Leibler divergence between the prior and the posterior distribution.

\begin{figure}
	\centerline{\includegraphics[width=0.70\columnwidth]{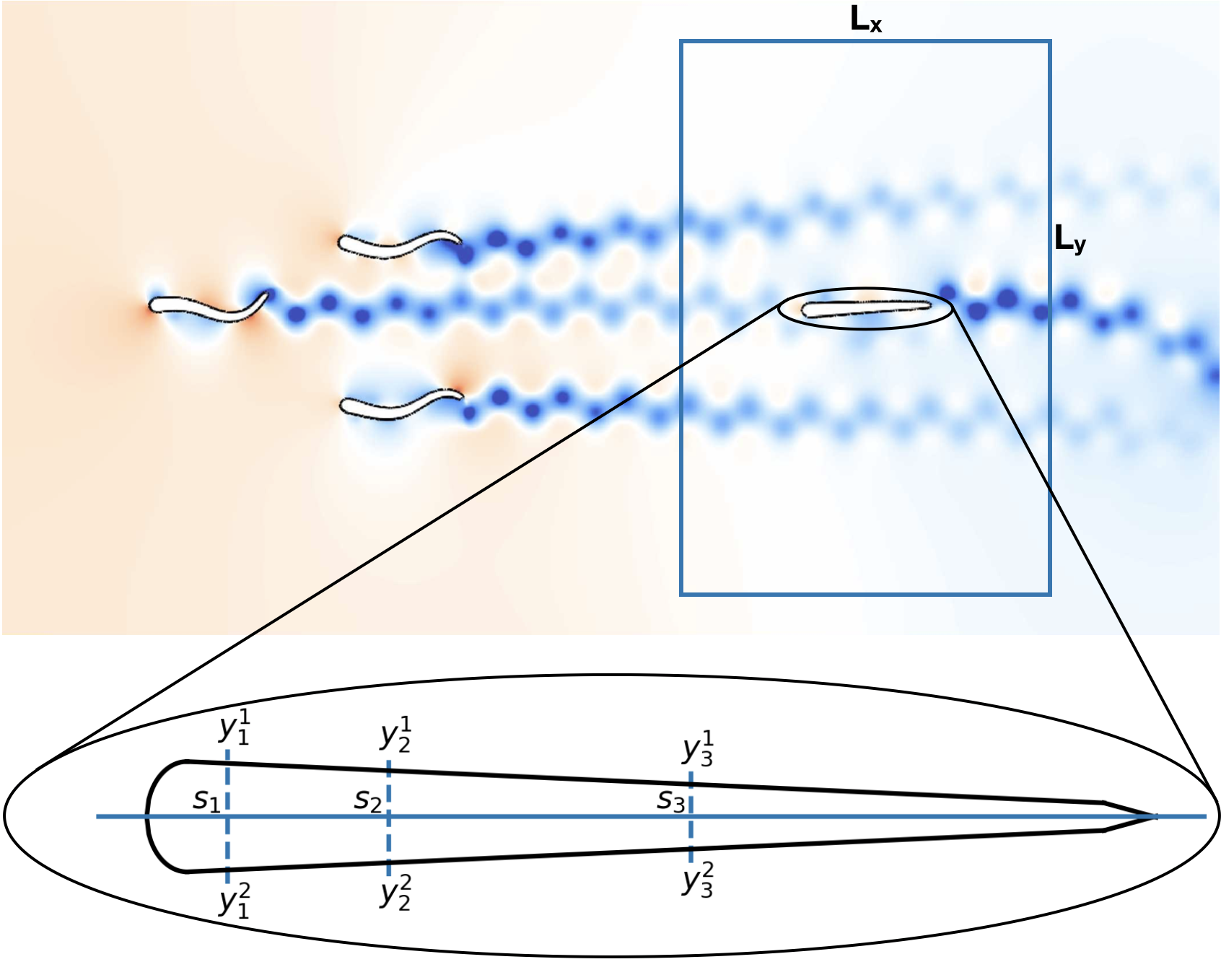}}
	\caption{
		Simulation setup used for determining the optimal sensor distribution on a fish-like body. The follower is initially located inside the rectangular area. The number of fish in the leading group is varied between one and eight. The sensor-placement algorithm attempts to find the arrangement of sensors $\SNS$ that allows the follower to determine with lowest uncertainty the relative position $\mathbf{r}$ and the number of fish $n_f$ in the leading group of fish. For each sensor $s_i$ the fish collects measurements $y_i^1$ and  $y_i^2$ at locations $\bm{x}_1(s_i)$ and $\bm{x}_2(s_i)$ on the skin, respectively.
	}
	\label{fig:setup}
\end{figure}

\subsubsection{Bayesian Estimation of Swimmers}

In the present  experiment setup, we consider a group of swimmers followed by a single swimmer. The follower needs to identify (i) the relative location $\mathbf{r}$ of the center of mass and (ii) the population $n_f$ of the leading group.
We denote with $\PRM = \mathbf{r}$ or $\PRM = n_f $ these unknown quantities and allow the follower  to update its prior belief $p(\PRM)$ about the leading group of fish by collecting measurements on its sensors. 
These sensors are distributed symmetrically on both sides of the fish and can be represented by a single point on its mid-line. We denote the $k$-th measurements location at the upper and the lower part with $\bm{x}_1(s_k)$ and $\bm{x}_2(s_k)$, respectively (see \cref{fig:setup} for  a sketch of the setup).

We denote by  $\Signal( \PRM; \SNS)\in{R}^{2n}$ the flow output  and  include an error term $\eps$ to account for inaccuracies such as as numerical errors and imperfections in the sensors. The  measurements on the fish body can be expressed as,

\begin{equation}
\MS = \Signal( \PRM; \SNS) + \eps \PERIOD
\label{eq:measurement}
\end{equation}
We model the error term by a multivariate Gaussian distribution $\eps\sim\mathcal{N}(\mathbf{0}, \boldsymbol{\Sigma}(\SNS))$ with zero mean and covariance matrix $\boldsymbol{\Sigma}(\SNS) \in \mathbb{R}^{2n \times 2n}$ so that the  likelihood of a measurement is expressed as,

\begin{equation}
\begin{split}
&p\left(\MS \vert \PRM, \SNS\right) = 
\dfrac{1}{\sqrt{(2\pi)^{2n} \det(\bm{\Sigma}(\SNS))}}\exp \left(-\dfrac{1}{2} 
\left(\MS - \Signal( \PRM; \SNS) \right)^T 
\bm{\Sigma}^{-1}(\SNS) 
\left(\MS - \Signal( \PRM; \SNS)  \right) \right) \PERIOD
\label{eq:likelihood}
\end{split}
\end{equation}
The covariance matrix  depends  on the sensor positions $\SNS$ and we assume that the prediction errors are correlated for measurements on the same side of the swimmer and uncorrelated if they originate from opposite sides. 
Finally, we assume that the correlation is decaying exponentially with the distance of the measurement locations. The functional form of the resulting covariance matrix is given by,

\begin{equation}
\Sigma_{ij}(\SNS) =
\begin{cases}
\sigma^2 \exp   \left( - \frac{ \| \bm{x}_1(s_i) - \bm{x}_1(s_j) \| }{\CorLen}\right),  & \quad \textrm{ if } 1\leq i,j \leq n \COMMA \\
\sigma^2 \exp   \left( - \frac{ \| \bm{x}_2(s_{i-n}) - \bm{x}_2(s_{j-n}) \| }{\CorLen}\right), & \quad \textrm{ if } n < i,j \leq 2n \COMMA \\
0   & \quad \textrm{ otherwise} \PERIOD
\end{cases}
\label{eq:covarianceMatrix}
\end{equation}
where $\CorLen>0$ is correlation length and $\sigma$ is the correlation strength. For all the cases described in this work, the correlation length is set to one tenth of the fish length $\CorLen=0.1L$. The correlation strength is set to be two times the mean over the signal coming from our simulations:

\begin{equation}
\sigma=\frac{1}{n\,N_{\PRM}} \sum_{j=1}^{2n} \sum_{i=1}^{N_{\PRM}}|\Signal(\PRM^{(i)};s_{j})| \PERIOD
\end{equation}
We remark that the covariance matrix must be symmetric and positive definite. In order to ensure that it is positive definite we have to handle the case where we try to pick a sensor location twice as in this case the positive definiteness is violated. We do so by setting the argument of the exponential to $10^{-7}$.
This form of the correlation error reduces the utility when sensors are placed too close together and prevents excessive clustering of the sensors~\cite{papadimitriou2012effect, simoen2013prediction}.


We wish to identify the locations $\SNS$ yielding the largest information gain about the unknown parameter $\PRM$ of the disturbance.
A measure for information gain is defined through the  Kullback-Leibler (KL) divergence between the prior belief of the parameter values and the posterior belief, i.e.\ after measuring the environment. 
The prior and posterior beliefs are represented through the density functions
$p(\PRM)$ and $p(\PRM | \MS,\SNS)$, respectively. We denote by $\PSPACE$ the support of $p(\PRM)$. The two densities are connected through Bayes' theorem,

\begin{equation}
p(\PRM | \MS,\SNS) = \frac{  p(\MS | \PRM  ,\SNS) \,  p(\PRM)  }{p(\MS)} \COMMA
\label{eq:bayes}
\end{equation}
where $p(\MS | \PRM  ,\SNS)$ is the likelihood function defined in \cref{eq:likelihood} and $p(\MS)$ is the normalization constant. 

The \emph{utility} function is defined as~\cite{Ryan2003},

\begin{equation}
u(\SNS,\MS) := D_{\text{KL}}(p(\PRM | \MS,\SNS)||p(\PRM))=\int_{\PSPACE}  \ln \frac{p(\PRM | \MS,\SNS)}{p(\PRM)}\; p( \PRM | \MS,\SNS) \dif \PRM \PERIOD
\label{eq:utilityFunction}
\end{equation}
The \emph{expected utility} is defined as the average value over all possible measurements,

\begin{equation}
\begin{split}
U(\SNS) :&= \mathbb{E}_{\MS | \SNS} \big [ u(\SNS,\MS) \big ] = \int_{\YSPACE} u(\SNS,\MS)\; p(\MS | \SNS) \dif \MS  \\
&=  \int_{\YSPACE}\int_{\PSPACE}  \; \ln \frac{p(\PRM | \MS,\SNS)}{p(\PRM)} \;p( \PRM | \MS,\SNS)\dif \PRM  \; p(\MS | \SNS)  \dif \MS 
\COMMA
\label{eq:expected:utility}
\end{split}
\end{equation}
where $\YSPACE$ is the support of measurements. Using \cref{eq:bayes} the expected utility can be expressed as,

\begin{equation}
\begin{split}
U(\SNS) = \int_{\YSPACE}  \int_{\PSPACE} \; \ln \frac{p(\MS | \PRM,\SNS)}{p(\MS | \SNS)} \;p( \MS |\PRM ,\SNS) \; p(\PRM) \dif \PRM  \dif \MS \PERIOD
\label{eq:utilityBayes}
\end{split}
\end{equation}
\subsubsection{Estimated expected utility for continuous random variables: school location}\label{sec:utilityContinuous}

When  $\PRM = \mathbf{r}$, the random variable takes values in a continuous domain.
The estimator for the expected utility in this case can be obtained by approximating the two integrals by Monte Carlo integration using $N_{\PRM}$ samples from $p(\PRM)$ and $N_{\MS}$ from $p( \MS |\PRM ,\SNS)$~\cite{huan2013simulation}. The resulting estimator is given by,

\begin{equation}
U(\SNS)\approx \hat U(\SNS) = \frac{1}{N_{\PRM} N_{\MS}} \sum_{j=1}^{N_{\MS}} \sum_{i=1}^{N_{\PRM}}  \left [   \ln p(\MS^{(i,j)} | 
\PRM^{(i)},\SNS) - \ln \left(  \frac{1}{N_{\PRM}} \sum_{k=1}^{N_{\PRM}} p(\MS^{(i,j)} | \PRM^{(k)},\SNS)  \right ) \right ]   \COMMA
\label{eq:estimator}
\end{equation}

where $\PRM^{(i)} \sim p_{\PRM}$ for $i=1,\ldots,N_{\PRM}$  and $\MS^{(i,j)} \sim \mathcal{N}( \Signal( \PRM^{(i)}; \SNS),\Sigma(\SNS))$ and $j=1,\ldots,N_{\MS}$.
We remark that the computational complexity of this procedure is mainly determined by the number of Navier-Stokes simulations $N_{\PRM}$. There is no additional computational burden to compute the $N_{\MS}$ samples following the measurement error model \cref{eq:measurement}.

\subsubsection{Estimated expected utility for discrete random variables: school size}\label{sec:utilityDiscrete}

When $\PRM$ is a discrete random variable with finite support taking values  in the set $ \{ \PRM_1,\dots,\PRM_{N_{\PRM}} \}$ the expected utility in \cref{eq:utilityBayes} is given by,

\begin{equation}
\begin{split}
    U(\SNS) =\sum\limits_{i=1}^{N_{\PRM}} p(\PRM_i)  \int_{\YSPACE} \; \ln \frac{p(\MS | \PRM_i,\SNS)}{p(\MS | \SNS)} \;p( \MS |\PRM_i ,\SNS) \; \dif \MS \PERIOD
\label{eq:expected:utility:discrete}
\end{split}
\end{equation}
Here, $\PRM = n_f$ represents the number of fish in the leading group. An estimator of the given utility can be obtained by Monte Carlo integration using $N_{\MS}$ samples from the likelihood distribution $p( \MS |\PRM_i ,\SNS)$. The estimator is given by

\begin{equation}
U(\SNS)\approx \hat U(\SNS) = \frac{1}{N_{\MS}} \sum_{j=1}^{N_{\MS}} \sum_{i=1}^{N_{\PRM}}  \, p(\PRM_{i}) \,   \left [   \ln p(\MS^{(i,j)} | 
\PRM_{i},\SNS) - \ln \left( \sum_{k=1}^{N_{\PRM}}  p(\PRM_{k}) p(\MS^{(i,j)} | \PRM_{k},\SNS)  \right ) \right ]   \PERIOD
\end{equation}
where $\MS^{(i,j)} \sim p( \MS |\PRM_i ,\SNS)$ for $j=1,\ldots,N_{\MS}$.
Let $\CFG$ be the random variable representing one of the group configurations.
Each group configuration is associated with a unique number $\CFG_{i,j}$ for $j=1,\ldots,n_i$, where $n_i$ is the total number of configurations containing $i$ fish. 
With this notation, $\CFG$ takes values in the set $\{\CFG_{i,j} \,|\,j=1,\ldots,n_i \, ,  i=1,\ldots,8  \}$. 
For an example of different configurations see \cref{app:configurations}
The likelihood can be written in terms of the variable $\CFG$ as

\begin{equation}
p(\MS | \PRM = \PRM_i,\SNS)=p(\MS | \CFG = \{ \CFG_{i,1}, \ldots, \CFG_{i,n_i}\} ,\SNS)  \PERIOD
\label{eq:likelihood:phi:a}
\end{equation}
Using the fact that for $i=1,\ldots,N_{\PRM}$,

\begin{equation*}
p(\MS, \CFG=\CFG_{k,\ell} | \PRM = \PRM_{i},\SNS) = 0, \,\, \textrm{ for } \,  k\neq i \COMMA
\end{equation*}
and

\begin{equation*}
p(\MS |  \PRM = \PRM_{i}, \CFG=\CFG_{i,\ell},\SNS) = p(\MS |, \CFG=\CFG_{i,\ell},\SNS),
\,\, \textrm{ for } \,  \ell=1,\ldots,n_i \COMMA 
\end{equation*}
and the assumption 

\begin{equation*}
p( \CFG=\CFG_{i,\ell} | \PRM = \PRM_{i}, \SNS ) = \frac{1}{n_i}, 
\,\, \textrm{ for } \,  \ell=1,\ldots,n_i \COMMA
\end{equation*}
the likelihood in \cref{eq:likelihood:phi:a} is written as,

\begin{equation}
\begin{split}
p(\MS | \PRM = \PRM_i,\SNS)
&= \sum_{k=1}^{N_{\PRM}}  \sum_{\ell=1}^{n_i}  p(\MS, \CFG=\CFG_{k,\ell} | \PRM = \PRM_{i},\SNS) \\
&=  \sum_{\ell=1}^{n_i}  p(\MS, \CFG=\CFG_{i,\ell} | \PRM = \PRM_{i},\SNS) \\
&=  \sum_{\ell=1}^{n_i}  p(\MS | \PRM = \PRM_{i},  \CFG=\CFG_{i,\ell} ,\SNS)  \, p( \CFG=\CFG_{i,\ell} | \PRM = \PRM_{i}) \\
&= \frac{1}{n_i} \sum_{\ell=1}^{n_i}  p(\MS |\CFG=\CFG_{i,\ell} ,\SNS) \PERIOD
\end{split}
\end{equation}
Notice that the likelihood is a mixture of distributions with equal weights and that $p(\MS |\CFG=\CFG_{i,\ell}) = \mathcal{N}( \MS | \Signal( \CFG_{i,\ell}; \SNS),\Sigma(\SNS)) $. In order to draw a sample from the likelihood, first we draw an integer $\ell^*$ with equal probability from 1 to 8 and them draw $y\MS\sim p(\MS |\CFG=\CFG_{i,\ell^*})$.

The final form of the estimator is given by

\begin{equation}
\begin{split}
\hat U(\SNS) = \frac{1}{N_{\MS}} \sum_{j=1}^{N_{\MS}} \sum_{i=1}^{N_{\PRM}}  \, 
p(\PRM_{i}) \, 
\left [ 
\ln \frac{1}{n_i} \sum_{\ell=1}^{n_i} p(\MS^{(i,j)} | \CFG_{i,\ell},\SNS) - \ln 
\left( 
\frac{1}{n_i} \sum_{k=1}^{N_{\PRM}} 
p(\PRM^{(k)})\sum_{\ell=1}^{n_i}  p(\MS^{(i,j)} | \CFG_{i,\ell},\SNS)   
\right ) 
\right ]   \PERIOD
\label{eq:estimator:discrete}
\end{split}
\end{equation}

\subsubsection{Optimization of the expected utility function}
\label{sec:sequentialOpt}

In order to determine the optimal sensor arrangement we maximize the utility estimator $\hat U(\SNS)$ described in \cref{eq:estimator}. 
It has been observed that the expected utility for many sensors often exhibit many local optima \cite{papadimitriou2012effect,papadimitriou:2015}.
Heuristic approaches, such as the sequential sensor placement algorithm described by \cite{Papadimitriou2004}, have been demonstrated to be effective alternatives.
Here, following \cite{Papadimitriou2004}, we perform the optimization iteratively, placing one sensor after the other. We start by placing one sensor $s_1^\star$ by a grid search in the interval $[0,0.9]$. In the next step we compute the location of the second sensor by setting $\SNS=(s_1^\star,s)$ and repeating the grid search for the new optimal location $s_2^\star$. This procedure is then continued by defining

\begin{equation}
s_i^\star = \argmax_{s}  \;  \hat{U}_i (\SNS) \qquad \textrm{where} \qquad \SNS=(s_1^\star,\ldots,s_{i-1}^\star,s) \PERIOD
\label{eq:optimal:sensor}
\end{equation}
We note that the scalar variable $s$ denotes the mid-line coordinate of a single sensor-pair, whereas the vector $\SNS$ holds the coordinate of all sensors-pairs. Besides the mentioned advantages, sequential placement allows to quantify the importance of each sensor placed and provides further insight into the resulting distribution of sensors.

\begin{figure}
	\centering
	\begin{subfigure}[b]{0.48\textwidth}
		\centering
		\includegraphics[width=\textwidth]{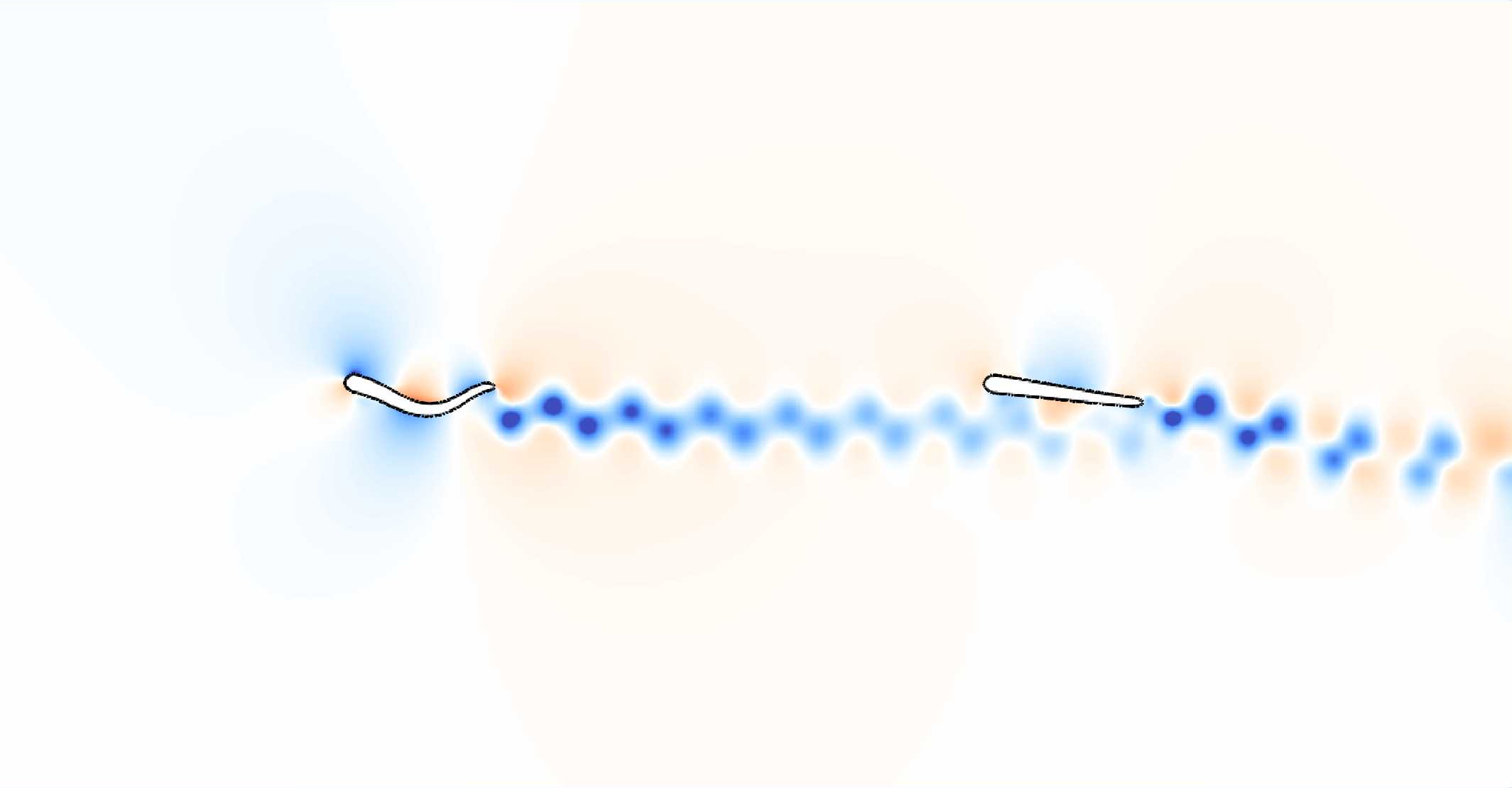}
		\subcaption{}
		\label{fig:snap1}
	\end{subfigure}
	\begin{subfigure}[b]{0.48\textwidth}
		\centering
		\includegraphics[width=\textwidth]{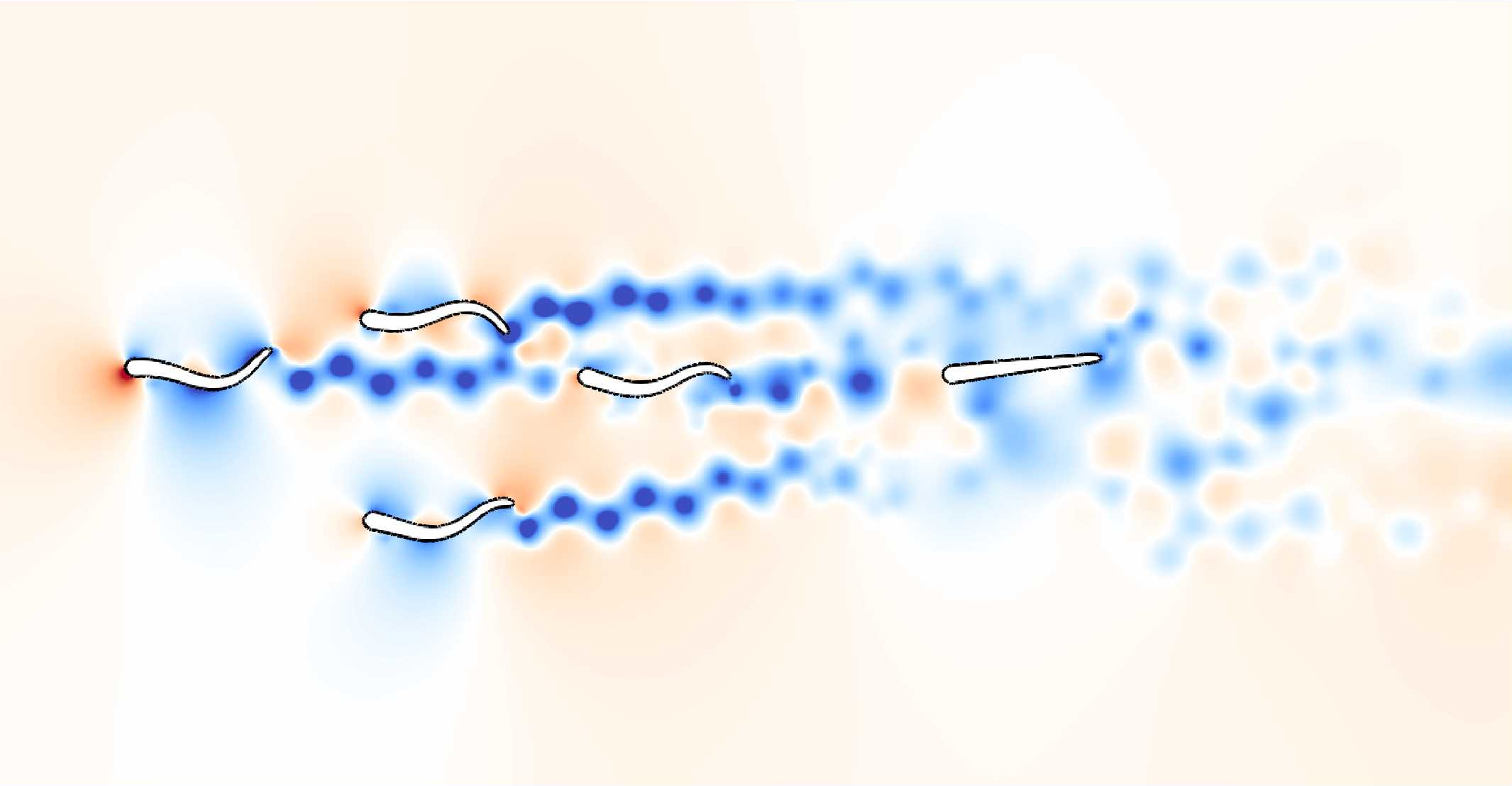}
		\subcaption{}
		\label{fig:snap4}
	\end{subfigure}
	\par\medskip
	\begin{subfigure}[b]{0.48\textwidth}
		\centering
		\includegraphics[width=\textwidth]{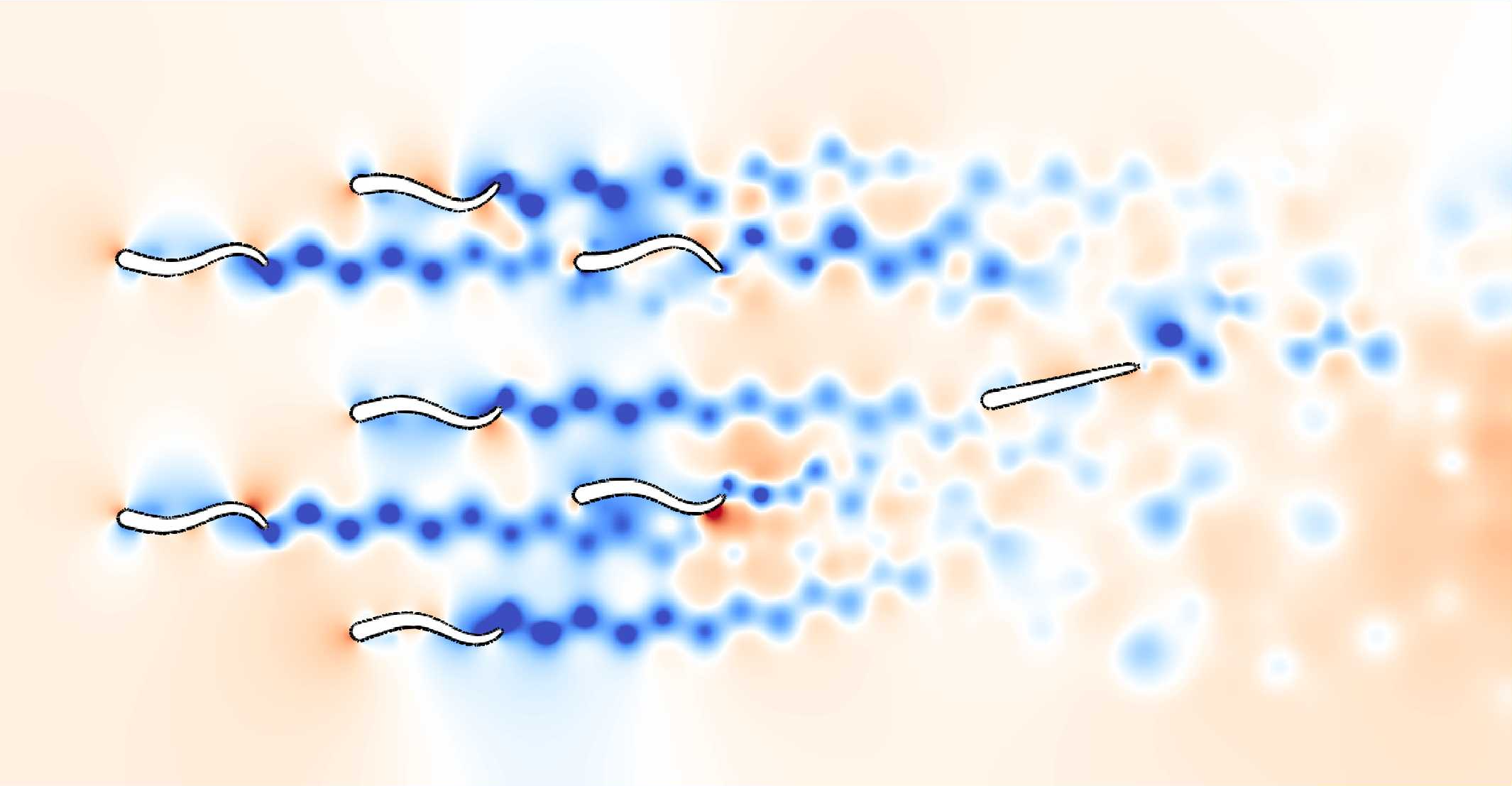}
		\subcaption{}
		\label{fig:snap7}
	\end{subfigure}
	\caption{
		Snapshots of the pressure field in the environment of the coasting swimmer generated by one (\cref{fig:snap1}), four (\cref{fig:snap4}) and seven (\cref{fig:snap7}) schooling swimmers. The snapshots are taken at the moment the measurement was performed for one particular location of the follower in the prior region. High pressure is shown in red and low pressure in blue.
		}
	\label{fig:snapshots}
\end{figure}

\section{Results}
\label{sec:results}

We examine the optimal arrangement of pressure gradient and shear stress sensors on the surface of a swimmer trailing a school of self-propelled swimmers. We consider two sensing objectives: (a) the size of the leading school and (b) the relative position of the school. 
The simulations correspond to a Reynolds number $\operatorname{Re}=\frac{L^2}{\nu}=2000$.
In all experiments we use $4096$ points to discretize the horizontal direction $x\in [0,1]$ and all artificial swimmers have a length of $L=0.1$.

For the ``\EXPa'' experiment the size of the group takes values $\PRM_i=1,\dots,8$. First we regard one configuration per group-size. In this case inferring the configuration is equivalent to inferring the number of fish in the group. To increase the difficulty we consider $n_i$ different initial configurations.
In each configuration we assign a number $\CFG_{i,\ell}$ for $i=1,\dots,8$ and $\ell=1,\dots,n_i$.
In total, we consider $N_{tot} = \sum_{i}n_i=61$ distinct configurations each having the same prior probability $1/N_{tot}$. 
In ~\cref{app:configurations} we present the initial condition for all configurations.
The center of mass of the school is located at $x=0.3$ and in the y-axis in the middle of the vertical extent of the domain.  
We use a controller to  fix the distance between $x$ and $y$ coordinates of two fish to $\Delta x=\Delta y = 0.15$, see \cref{sec:formation}. 

For the ``\EXPb'' experiment we consider three independent experiments with one, four and seven leading fish. 
Snapshots of the pressure field for these simulations are presented in \cref{fig:snapshots}.
The prior probability for the position of the group is uniform in the domain $[0.6,0.8]\times [0.1,0.4]$. 
The support of the prior probability is discretized with $21\times 31$ gridpoints.
Since the experiments are independent, the total expected utility function for the three cases is the sum of the expected utility of each experiment \cite{Verma2019}.

For both experiments we record the pressure gradient and shear stress on the surface of the fish using the methods discussed in~\cref{sec:sensors}. 
The motion of the fish introduces disturbances on it's own surface. 
In order to distinguish the self-induced from the environment disturbances we freeze the movement of the following fish and set its curvature to zero. 
The freezing time is selected by evolving the simulation until the wakes of the leading group are sufficiently mixed and passed the following swimmer. 
We found that this is the case for $T=22$. 
The transition from swimming to coasting motion takes place during the time interval $[T,T+1]$. 
Finally, we record the pressure gradient and the shear stress at time $T+2$.
The resulting sensor-signal associated to the midline coordinates $\SNS$ for a given configuration $\PRM$ is denoted $\Signal(\PRM;\SNS)$, see \cref{eq:measurement}.

\subsection{Utility function for the first sensor} \label{sec:results:single:sensos}

\begin{figure}
	\centering
	\begin{subfigure}[c]{0.48\textwidth}
		\centering
		\includegraphics[width=\textwidth]{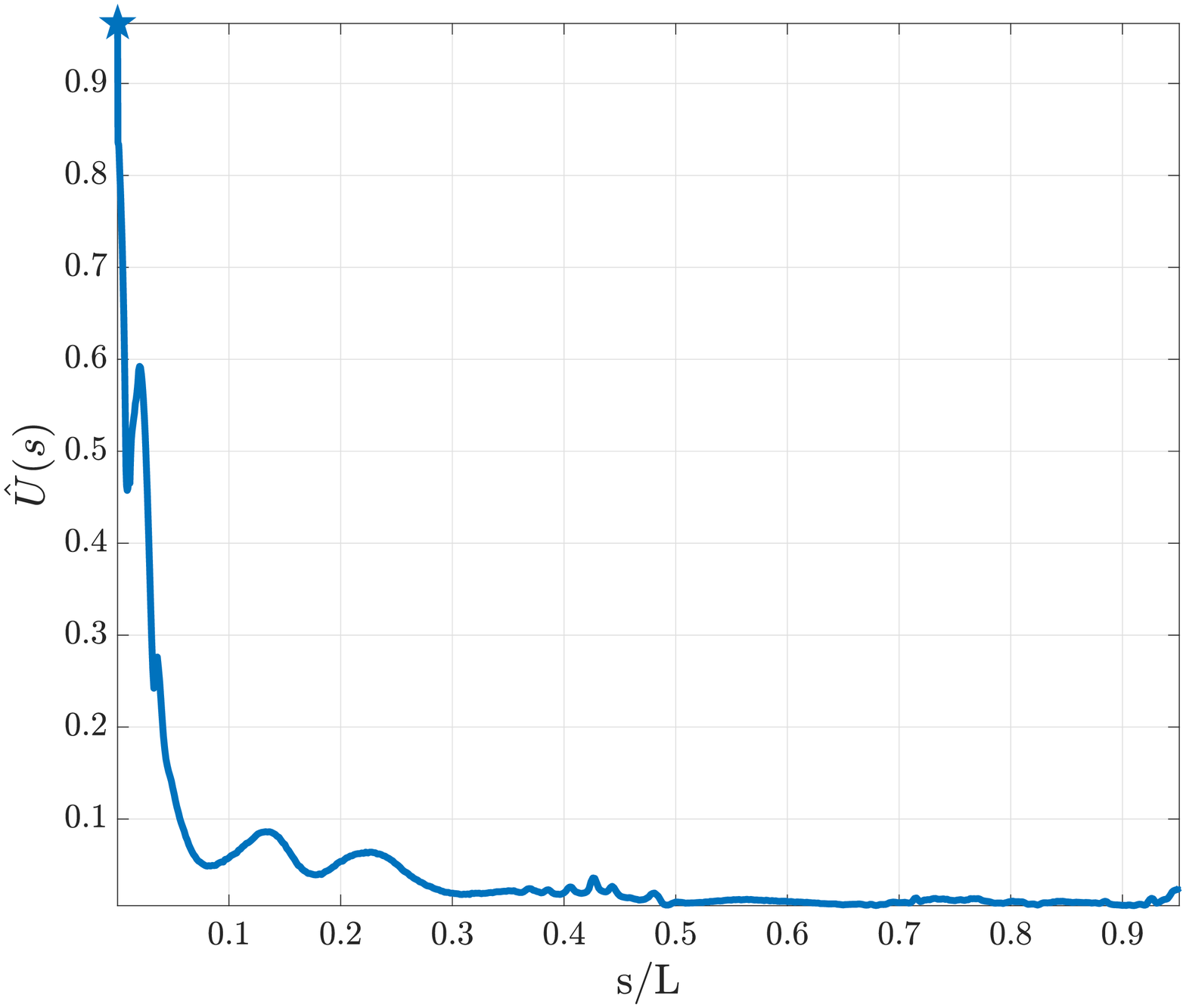}
		\subcaption{}
		\label{fig:numFishUtility}
	\end{subfigure}
	\begin{subfigure}[c]{0.48\textwidth}
		\centering
		\includegraphics[width=\textwidth]{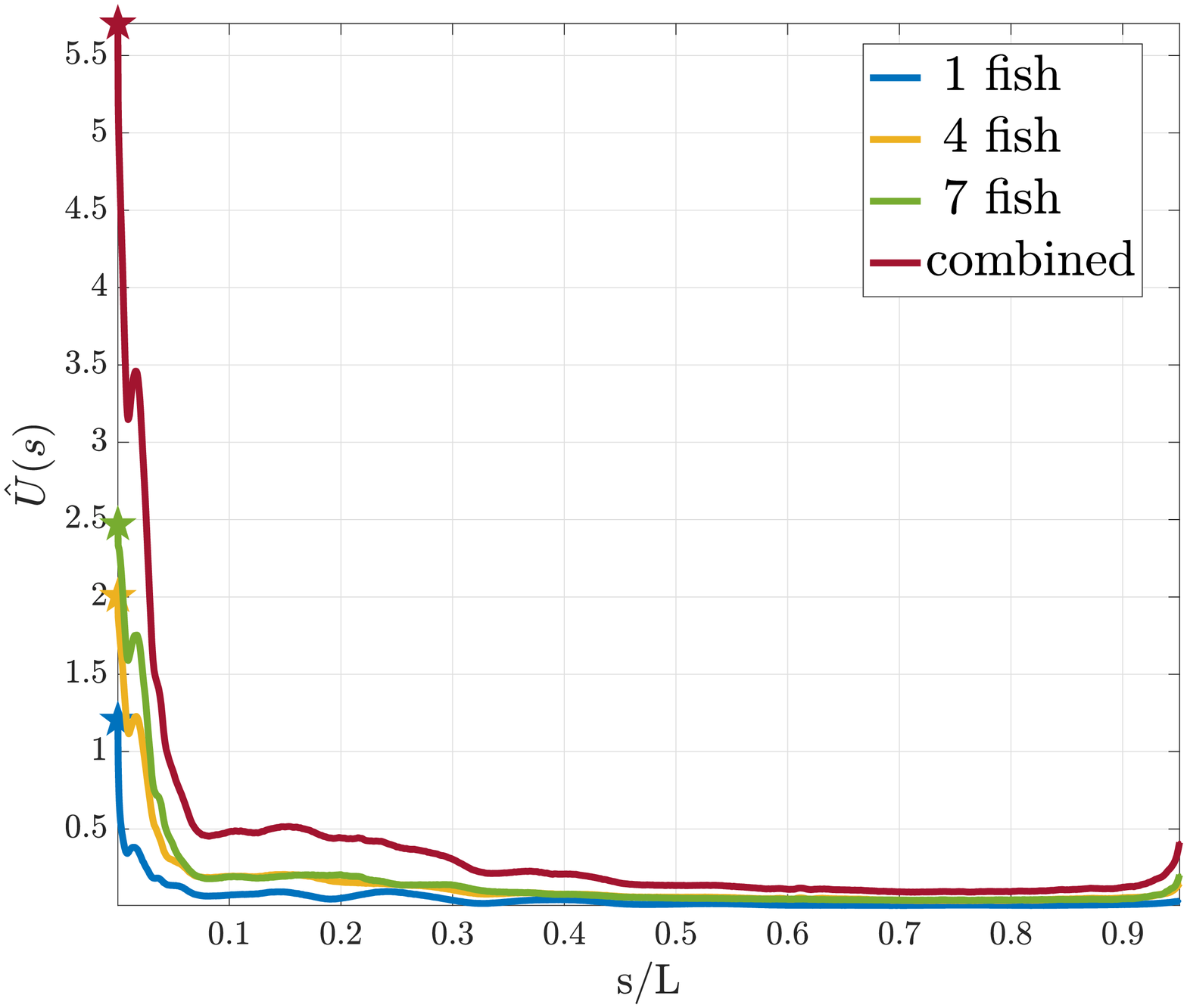}
		\subcaption{}
		\label{fig:locFishUtility}
	\end{subfigure}
	\caption{
Utility curves for the first sensor using pressure measurements. In \cref{fig:numFishUtility} the utility estimator for the ``\EXPa'' experiment is presented. \Cref{fig:locFishUtility} corresponds to the utility estimator for the ``\EXPb'' experiment.  We show the resulting curves for one, three and seven fish in the leading group and the total expected utility. We observe that although the form does not drastically change, the total utility increases with increasing size of the leading group.}
	\label{fig:one:sensor:utility}
\end{figure}

In this section we discuss the optimal location of a single pressure gradient sensor using the estimators in \cref{eq:estimator} and \cref{eq:estimator:discrete}. 
Higher values of the expected utility correspond to preferable locations for the sensor since the information gain at these sensors is higher. 
The resulting utilities are plotted in~\cref{fig:one:sensor:utility}. For all experiments we find that the tip of the head ($s=0$) exhibits the largest utility independent of the number of fish in the leading group.

At the tip of the head the two symmetrically placed sensors have the smallest distance.
In \cref{eq:covarianceMatrix} we have assumed that the two fish-halves are symmetric and uncorrelated.
Due to the small distance of the sensors at the head, spatial correlation between the sensors across the fish-halves would decrease the utility of this location.
In order to test whether the utility for sensors at the head is influenced by this symmetry assumption, we perform experiments where we place a single sensor on one side of the fish.
Again in this case the location at the head is found to have the highest expected utility.

Based on the evidence that the head experiences the largest variance of pressure gradients $\Signal( \PRM; \SNS)$ and that this is correlated with the density of sub-canal neuromast~\cite{Ristoph2015}, we examine the variance of the values obtained from our solution of the Navier-Stokes equation. 
We confirm that this observation is consistent with our simulations. We find that independent of the number of fish, the variance in the sensor signal $\operatorname{var}_{\PRM}(\Signal(\PRM;\SNS))$ is largest at $s=0$.

\subsection{Sequential sensor placement}\label{sec:results:many:sensors}

\begin{figure}
	\centering
	\begin{subfigure}[c]{0.49\textwidth}
		\centering
		\includegraphics[width=\textwidth]{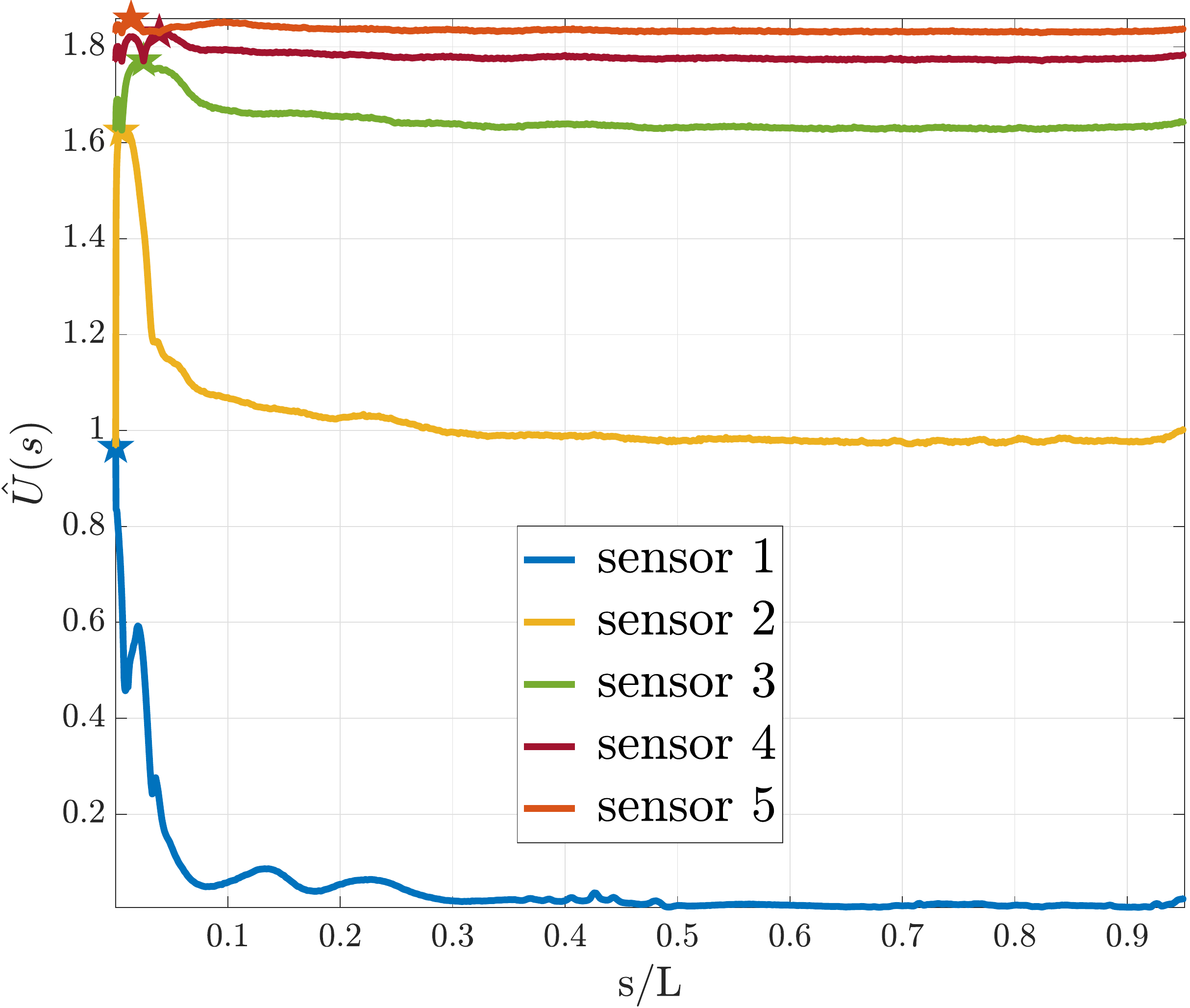}
		\subcaption{}
		\label{fig:utilityNumFish}
	\end{subfigure}
	\begin{subfigure}[c]{0.49\textwidth}
		\centering
		\includegraphics[width=\textwidth]{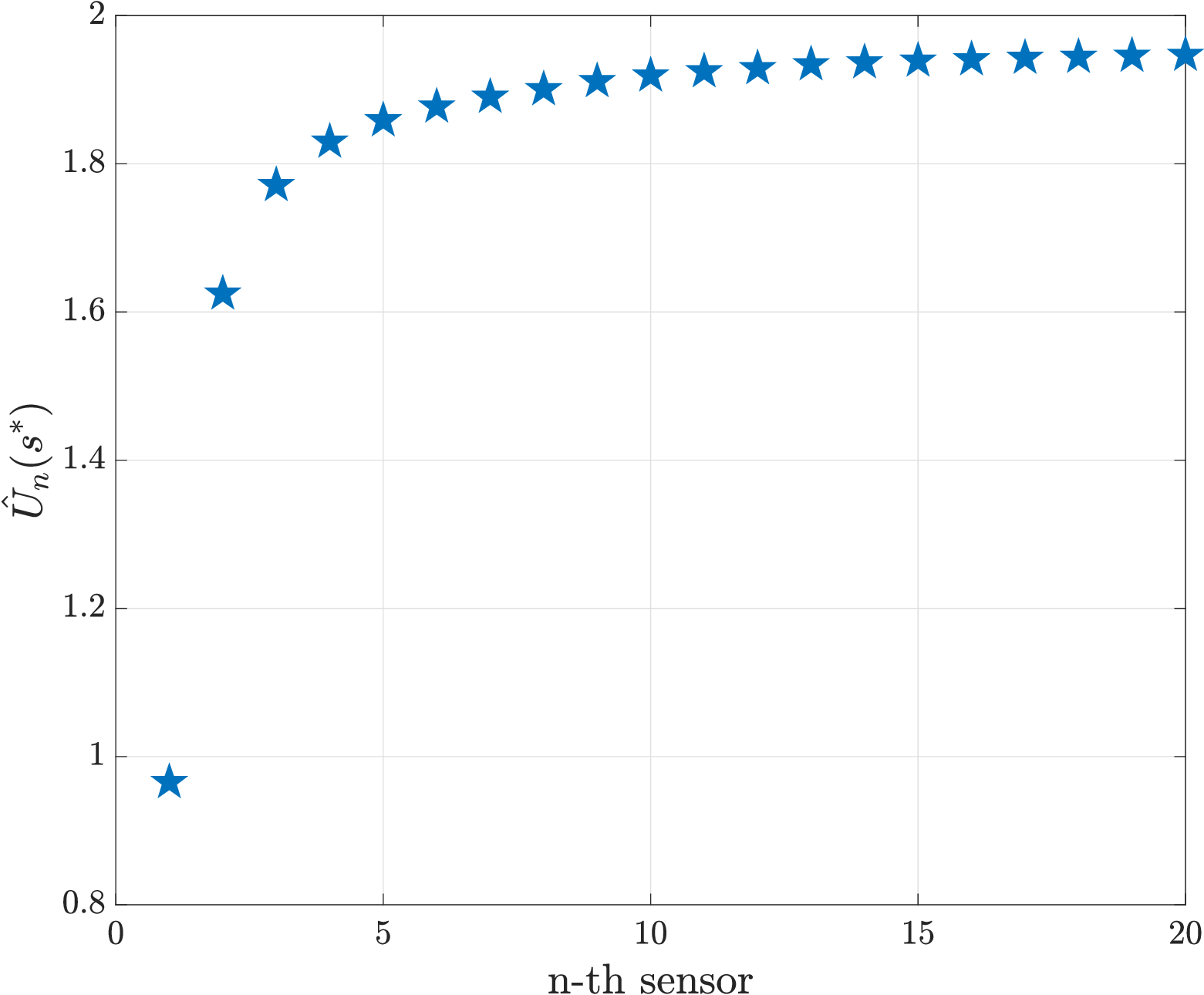}
		\subcaption{}
		\label{fig:numFishConvergence}
	\end{subfigure}
	\par\medskip
	\begin{subfigure}[c]{0.9\textwidth}
		\centering
		\includegraphics[width=\textwidth]{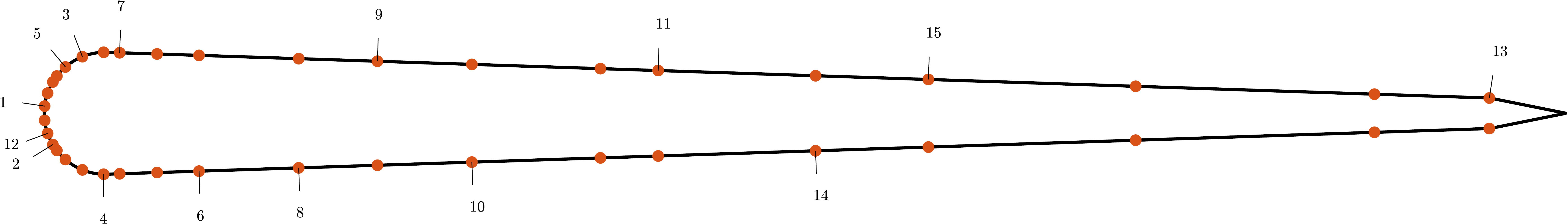}
		\subcaption{}
		\label{fig:numFishSensors}
	\end{subfigure}
	\caption{Optimal sensor placement for the pressure sensors and the ``\EXPa'' experiment. In \cref{fig:utilityLocation} the utility estimator for the first five sensors and in \cref{fig:locationCovergence} the value of the utility estimator at the optimal sensor location for the first $20$ sensors are presented. In \cref{fig:locationSensors} the distribution of  the sensors on the fish surface is presented.}
	\label{fig:sequentialPlacementNumFish}
\end{figure}

In this section we discuss the results of the sequential sensor placement described in \cref{sec:sequentialOpt}.  
For the ``\EXPa'' experiment we present the results in \cref{fig:sequentialPlacementNumFish}. 
In \cref{fig:utilityNumFish} the utility curve for the first five sensors is shown. 
We observe that the utility curve becomes flatter as the number of sensors increase. 
Furthermore, we observe that the location where the previous sensor was placed is a minimum for the utility for the next sensor. 
\Cref{fig:numFishConvergence} shows the utility estimator at the optimal sensor for up to $20$ sensors and it is evident that the value of the expected utility reaches a plateau. 
In~\cref{fig:numFishSensors} the location of the sensors on the skin if the fish is presented. 
The numbers correspond to the iteration in the sequential procedure that the sensor was placed.
Note that the sensors are being placed symmetrically. 

\begin{figure}
	\centering
	\begin{subfigure}[c]{0.49\textwidth}
		\centering
		\includegraphics[width=\textwidth]{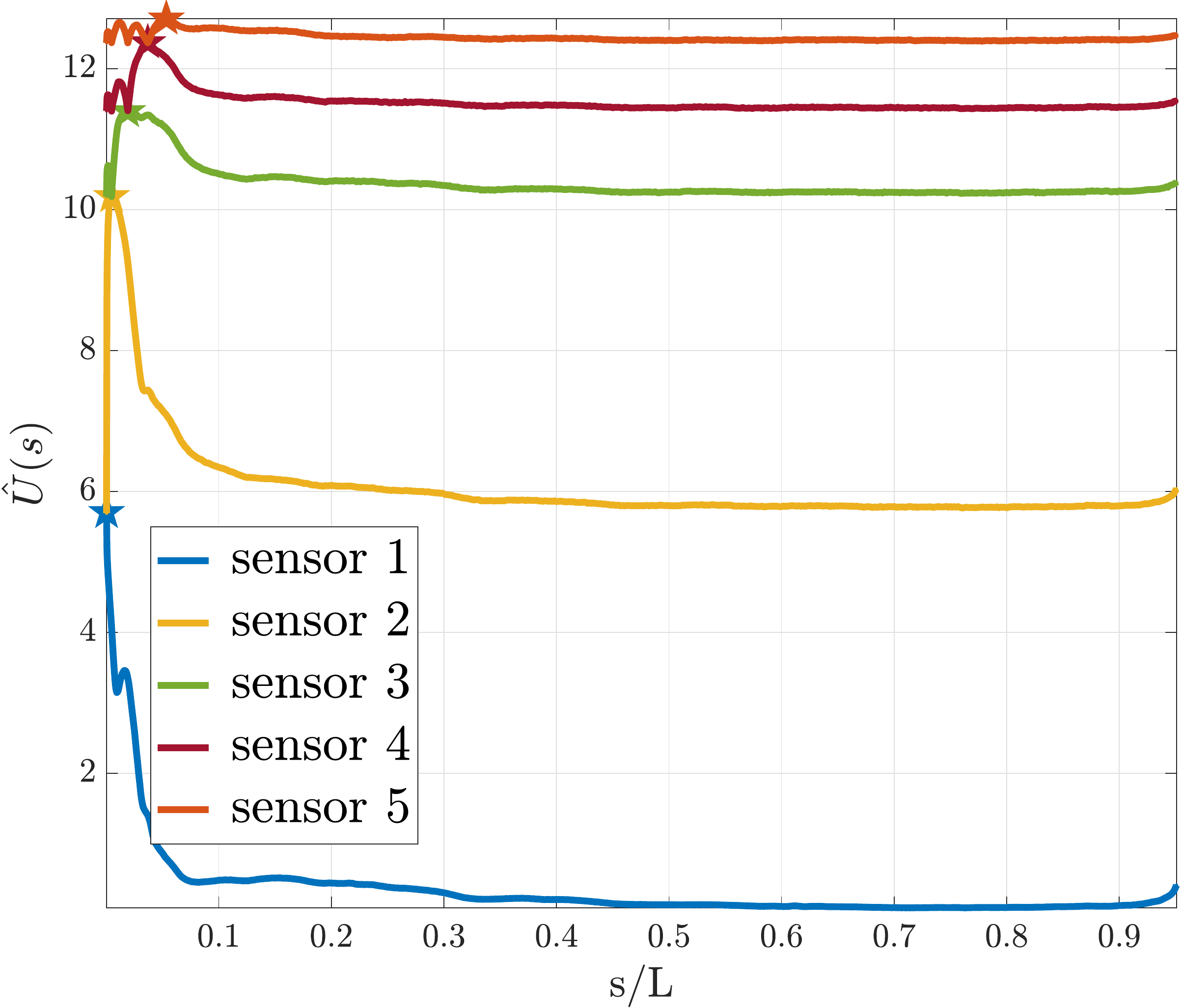}
		\subcaption{}
		\label{fig:utilityLocation}
	\end{subfigure}
   \begin{subfigure}[c]{0.49\textwidth}
   	\centering
   	\includegraphics[width=\textwidth]{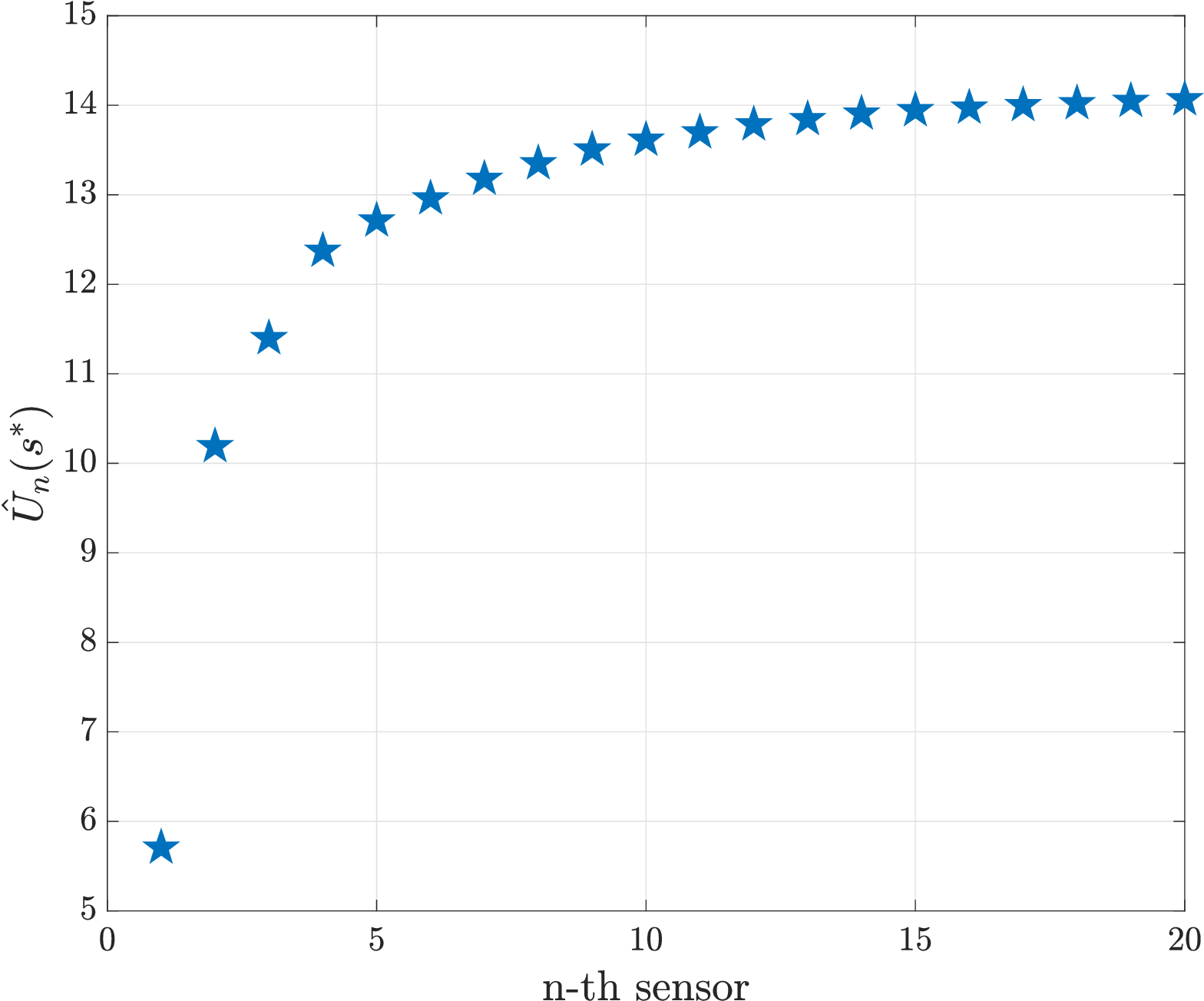}
   	\subcaption{}
   	\label{fig:locationCovergence}
   \end{subfigure}
		\par\medskip
	\begin{subfigure}[c]{0.9\textwidth}
		\centering
		\includegraphics[width=\textwidth]{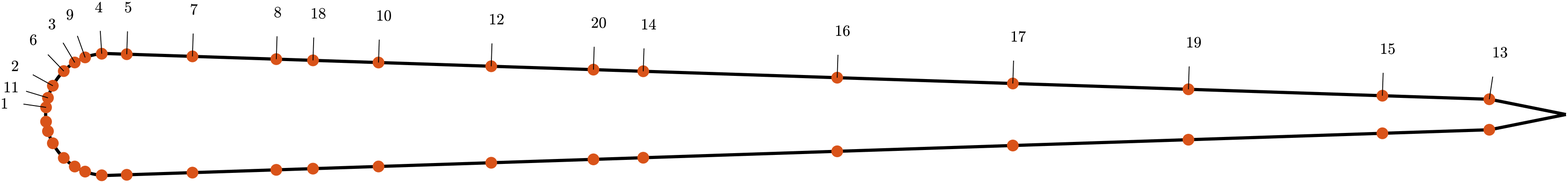}
		\subcaption{}
		\label{fig:locationSensors}
	\end{subfigure}
	\caption{Optimal sensor placement for the pressure gradient sensors for the ``\EXPb'' experiment. In \cref{fig:utilityLocation} the utility estimator for the first five sensors and in \cref{fig:locationCovergence} the value of the utility estimator at the optimal sensor location for the first $20$ sensors are presented. In \cref{fig:locationSensors} the distribution of  the sensors on the fish surface is presented.}
	\label{fig:sequentialPlacementLocation}
\end{figure}

The optimal sensor placement results for the ``\EXPb'' experiment  can be found in~\cref{fig:sequentialPlacementLocation}. Similar to the other experiment the utility curves become flatter after every placed sensor  and the location for the previous sensor is a minimum for the utility for the next sensor (see \cref{fig:utilityLocation}). We plot the maximum of the utility for up to $20$ sensors (see \cref{fig:locationCovergence}) and observe a convergence to a constant value. 
In \cref{fig:locationSensors} the location of the first 20 sensors is presented.

For both experiments it is evident that the utility of the optimal sensor location approaches a constant value. 
This fact can be explained by recalling that the expected utility \cref{eq:utilityBayes} is a measure of the averaged distance between the prior and the posterior distribution. Increasing the number of sensors leads to an increase in the number of measurements. By the Bayesian central limit theorem, increasing the number of measurements leads to convergence of the posterior to a Dirac distribution. As soon as the posterior has converged, the expected distance from the prior, and thus the expected utility, remains constant.

\subsection{Inference of the environment}
\label{sec:results:inference}

In this section we demonstrate the importance of the optimal sensor locations and examine the convergence of the posterior distribution. 
We compute the posterior distribution via Bayes rule given in \cref{eq:bayes}. 
We set $\MS=\mathbf{F}(\PRM, \SNS)$ and compute the posterior for different values of $\PRM$ in the prior region. 
We consider measurements collected at: (a)  the optimal   and (b) the worst sensors location. 

\begin{figure}
	\centering
	\begin{subfigure}[b]{0.48\textwidth}
		\centering
		\includegraphics[width=\textwidth]{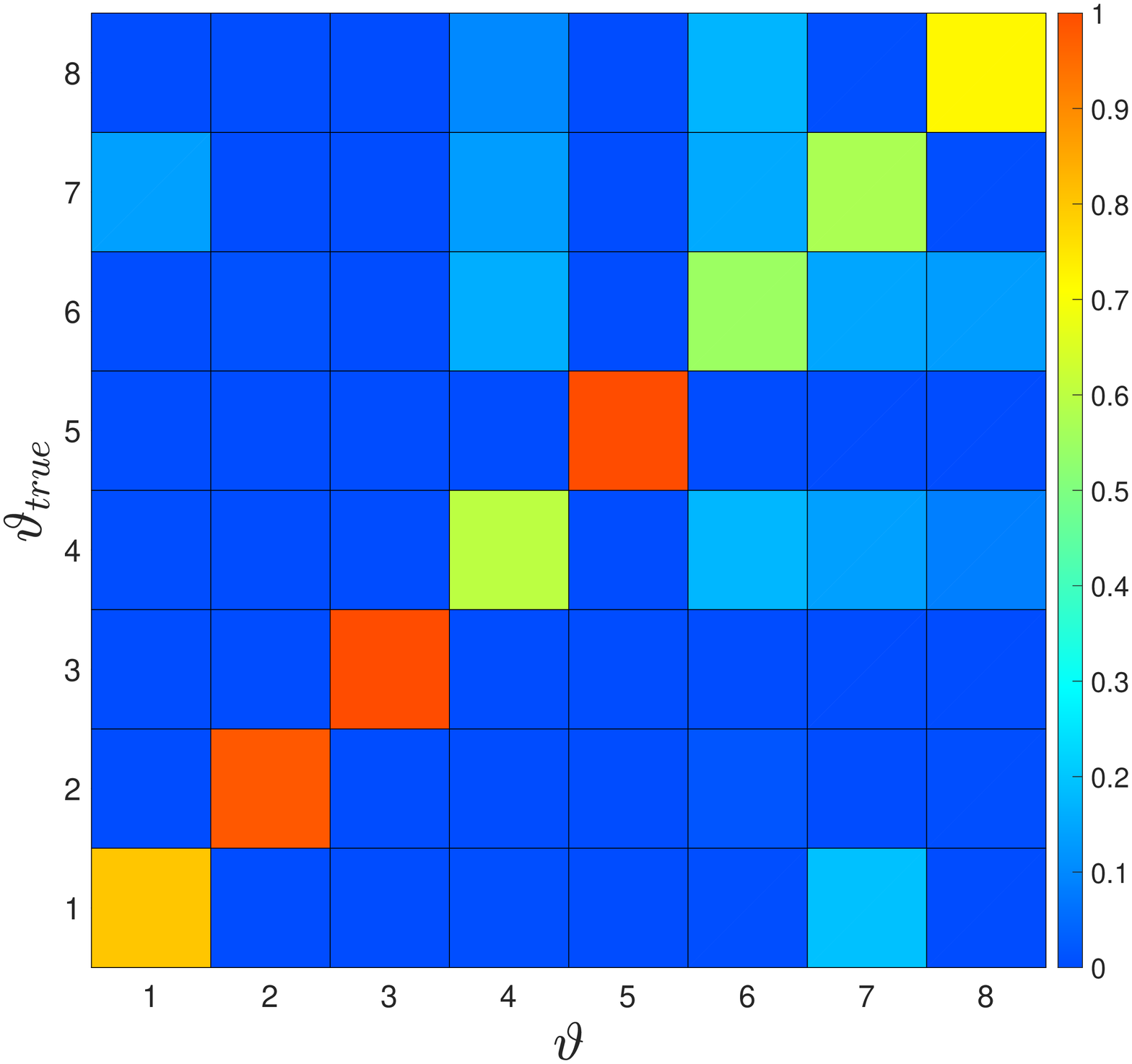}
		\subcaption{}
		\label{fig:barplotOSP-simple}
	\end{subfigure}
	\hspace{5pt}
	\begin{subfigure}[b]{0.48\textwidth}
		\centering
		\includegraphics[width=\textwidth]{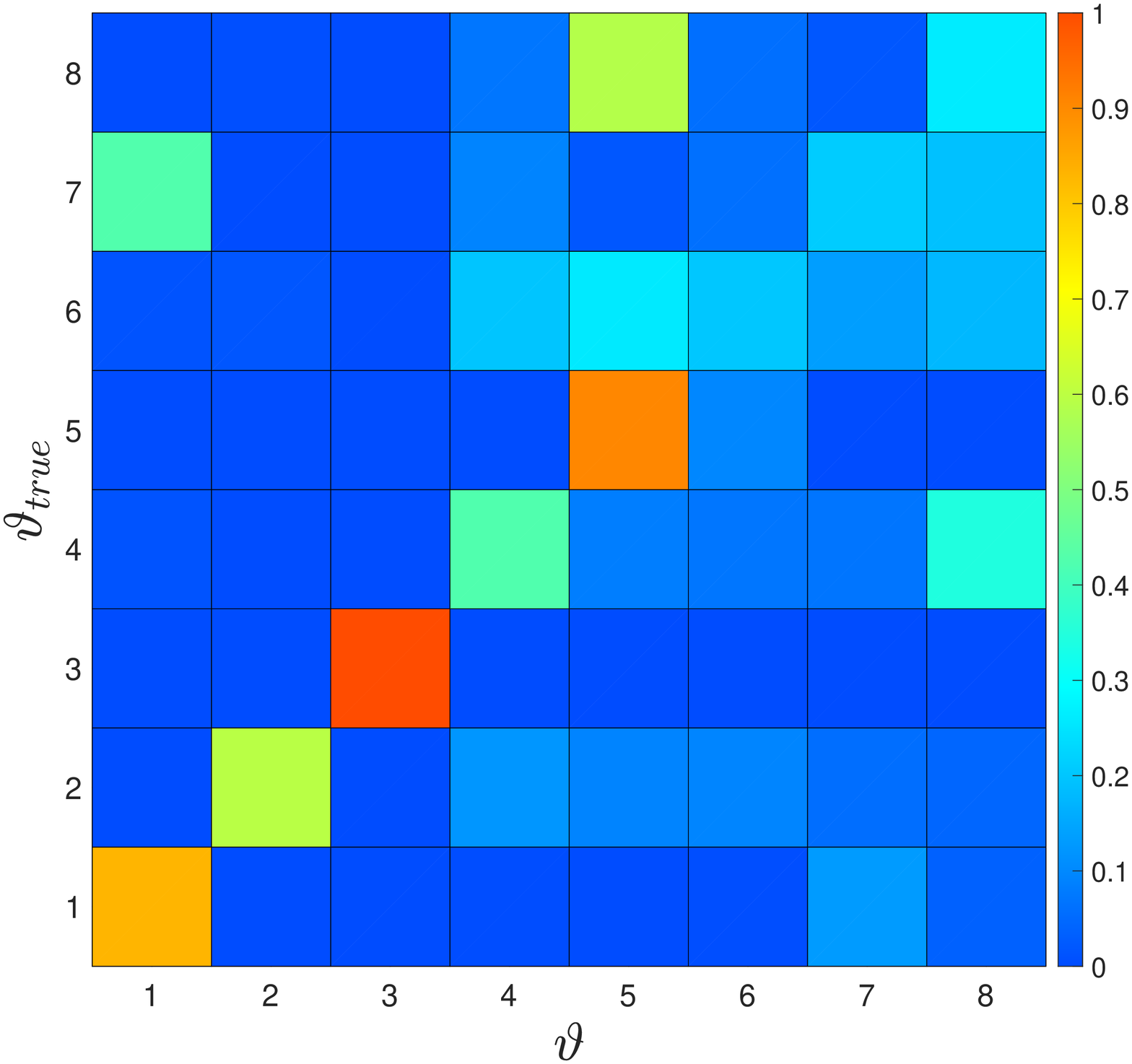}
		\subcaption{}
		\label{fig:barplotOSP}
	\end{subfigure}
	\hspace{5pt}
		\vspace{10pt}
	\begin{subfigure}[b]{0.48\textwidth}
		\centering
		\includegraphics[width=\textwidth]{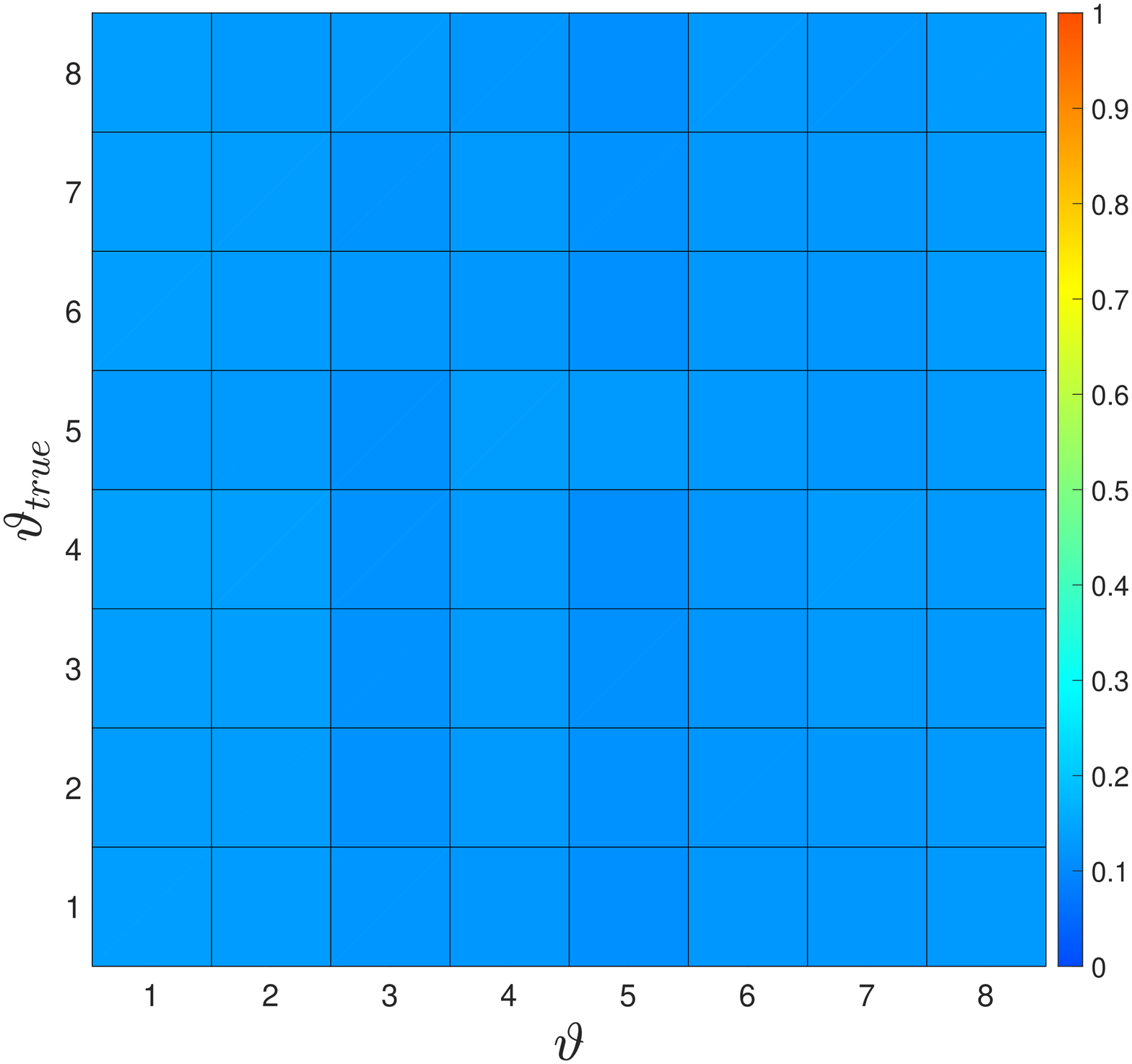}
		\subcaption{}
		\label{fig:barplotWSP}
	\end{subfigure}
	\caption{
        (a) Estimated posterior probability for a single sensor optimally placed and a single configuration per group size. 
        The posterior shows clear peaks at the correct number of fish for all cases, leading to perfect inference of the parameter of inference.
        The posterior probability for (b) optimal and (c) worst sensor location for multiple configurations per group size.
        Here, for the optimal sensor location for one, two, three and five fish shows a clear peak for the true size of the group. For the worst sensor location the posterior is almost uniform and does not allow to extract any information about the size of group.}
	\label{fig:NumFishPosterior}
\end{figure}

The posterior probability for the ``\EXPa'' experiment is shown in \cref{fig:NumFishPosterior}. We observe that the worst sensor location implies an almost uniform posterior distribution, reflecting  that measurements at this sensor carry no information. On the other hand, the posterior distribution for the optimal sensor is more informative. We observe that for groups with small size the follower is able to identify the size with more confidence, as opposed to larger groups. We compare the posterior for an experiment with only one configuration per group-size to an experiment with multiple configurations. For multiple configurations the posterior is less informative. This indicates that the second case occurs to be a more difficult problem. Finally, notice that the posterior for one configuration is symmetric, where when adding multiple configurations this symmetry is broken. This fact is discussed in \cref{app:posterior:symmetry}.

\begin{figure}
	\centering
	\begin{subfigure}[b]{0.49\textwidth}
		\centering
		\includegraphics[width=\textwidth]{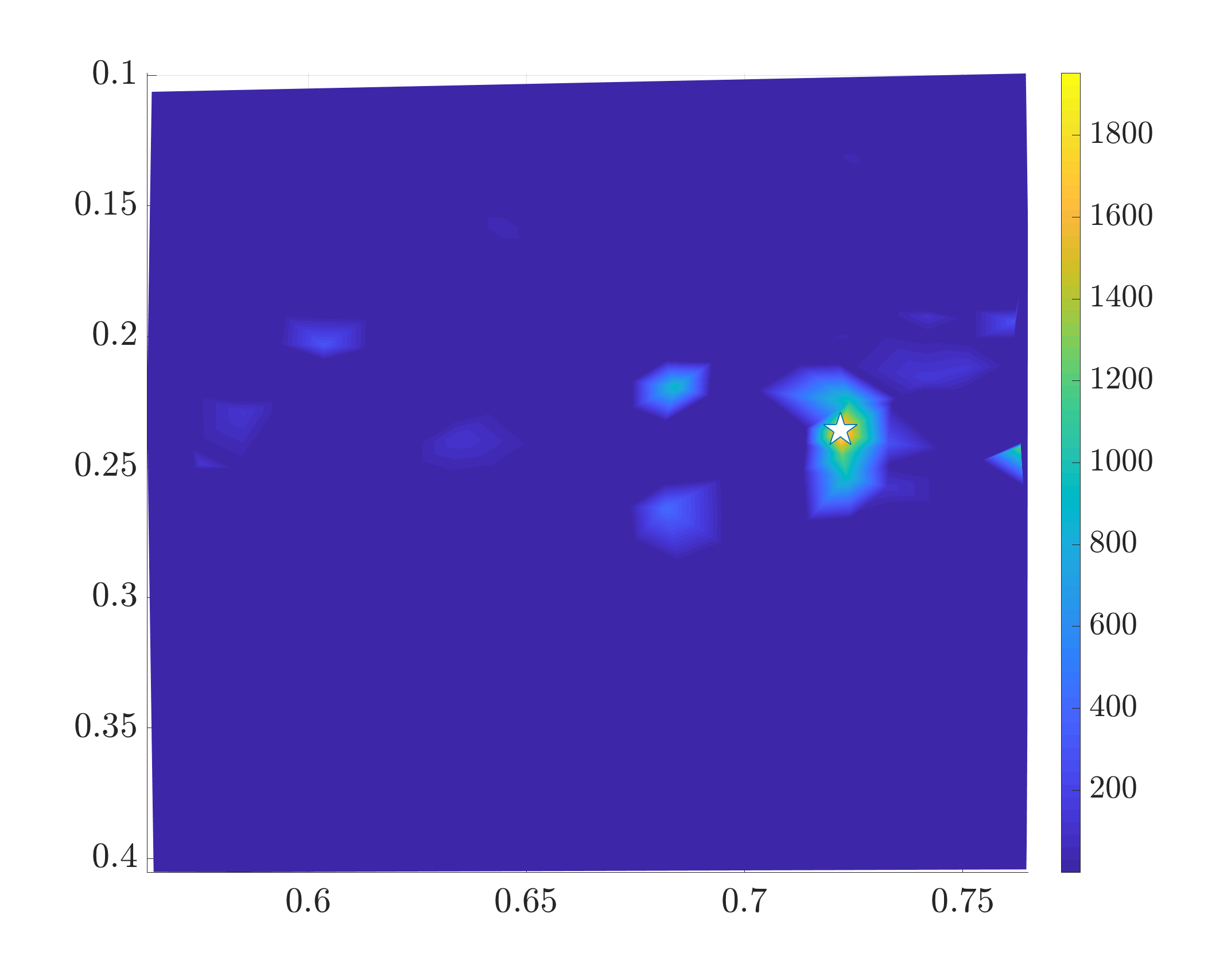}
		\vspace{-20pt}
		\subcaption{One sensor, best location}
		\label{fig:posterior1sensorBest}
	\end{subfigure}
	\vspace{10pt}
	\begin{subfigure}[b]{0.49\textwidth}
		\centering
		\includegraphics[width=\textwidth]{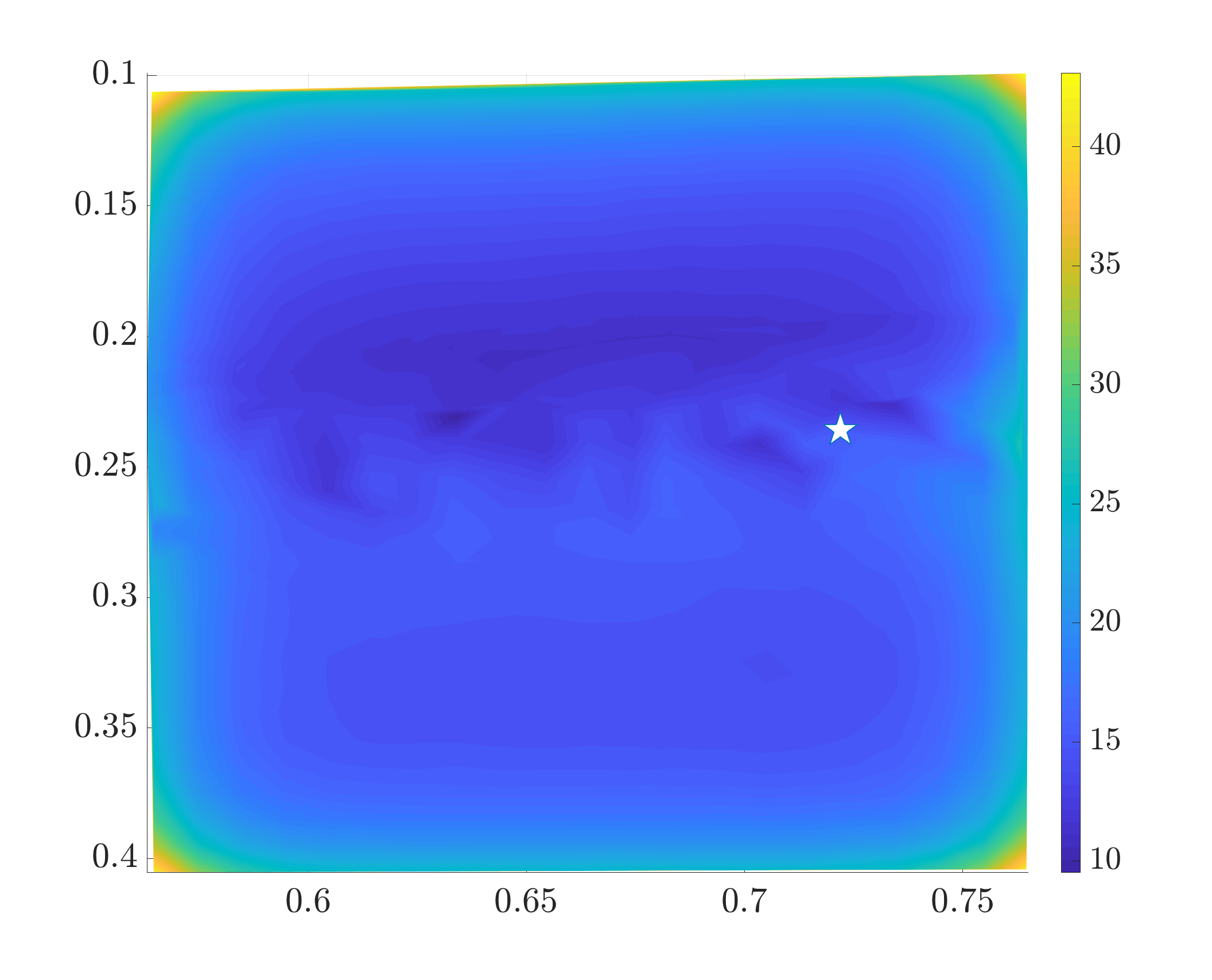}
		\vspace{-20pt}
		\subcaption{One sensor, worst location}
		\label{fig:posterior1sensorWorst}
	\end{subfigure}
	\vspace{10pt}
\begin{subfigure}[b]{0.49\textwidth}
	\centering
		\includegraphics[width=\textwidth]{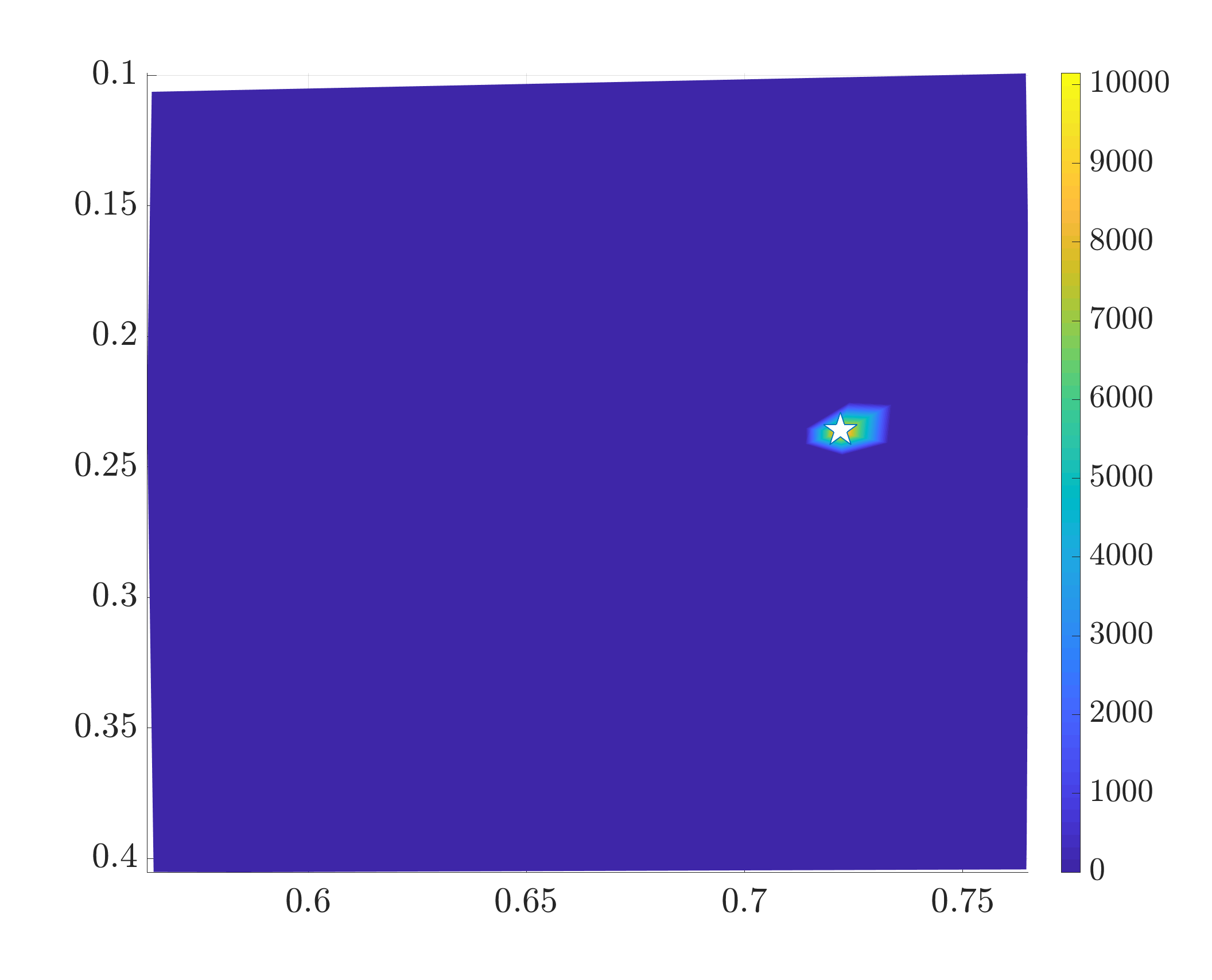}
	\vspace{-20pt}	
	\subcaption{Three sensors, best location}
	\label{fig:posterior3sensorBest}
\end{subfigure}
\begin{subfigure}[b]{0.49\textwidth}
	\centering
		\includegraphics[width=\textwidth]{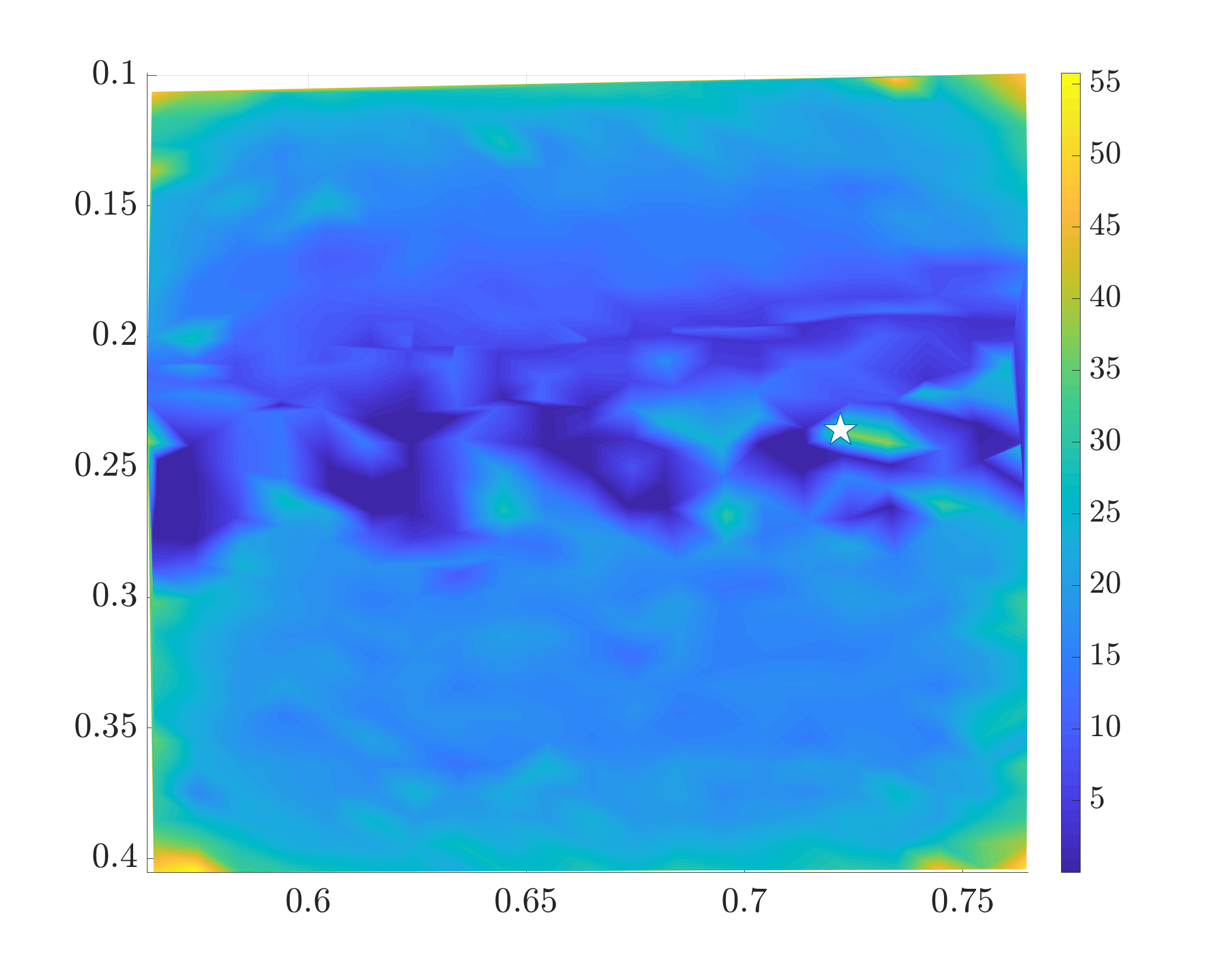}
	\vspace{-20pt}
	\subcaption{Three sensors, worst location}
	\label{fig:posterior3sensorWorst}
\end{subfigure}
    \vspace{10pt}
	\caption{
		Estimated posterior for the final location for the best (left column) and worst (right column) sensor-location for one (upper row) and three sensors (lower row). Light colors correspond to high probability density values. We marked the actual location with a white star.
	}
	\label{fig:LocationPosterior}
\end{figure}

The posterior density for the ``\EXPb'' with one leading fish is presented in \cref{fig:LocationPosterior}.
The posterior for the configuration with three and seven fish is similar.
We compute the posterior for measurements at the best and the worst location for one and three sensors.
For the three sensors the worst location has been selected in all three phases of the sequential placement.
The results for the normalized densities are shown in \cref{fig:LocationPosterior}.
We observe that one sensor at the optimal location gives a very peaked posterior. Three optimal sensors can infer the location with low uncertainty. This is not the case for the worst sensors, where adding more sensors does not immediately lead to uncertainty reduction.

\subsection{Shear stress sensors} \label{sec:results:shear}

In this section we discuss the results for the optimal positioning of shear stress sensors. 
We follow the same procedure as in \cref{sec:results:single:sensos} and \cref{sec:results:many:sensors}.
Here, we omit the presentation of the results and  focus on the similarities and differences to the pressure gradient sensors.

The optimal location for a single sensor for the ``\EXPa'' experiment is at $s^*=3.01\cdot 10^{-4}$. For the ``\EXPb'' experiment we find the optimal location $s^*=3.84\cdot 10^{-4}$. In contrast to the optimal location for one pressure gradient sensor, the found sensor is not at the tip of the head and is at different positions for the two experiments. Examining the variance in the shear signal shows quantitatively the same behaviour as the utility. Comparing the location of the maxima shows that they do not coincide for shear sensors.

\begin{figure}
	\centering
	\begin{subfigure}[c]{0.9\textwidth}
		\centering
		\includegraphics[width=\textwidth]{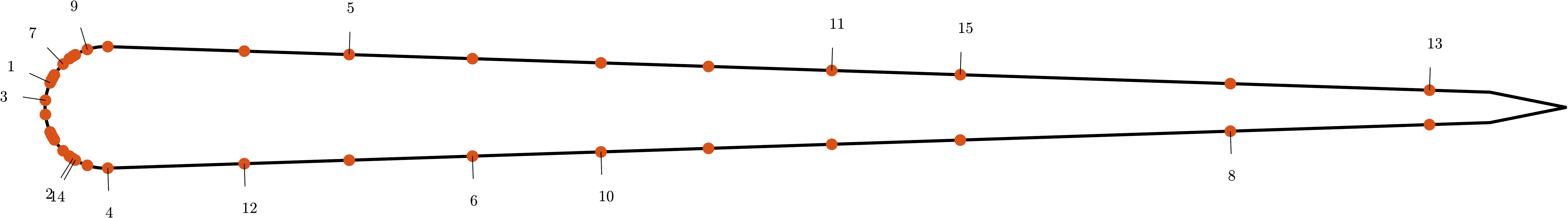}
		\subcaption{}
		\label{fig:numFishShearSensors}
	\end{subfigure}
	\par\medskip
	\begin{subfigure}[c]{0.9\textwidth}
		\centering
		\includegraphics[width=\textwidth]{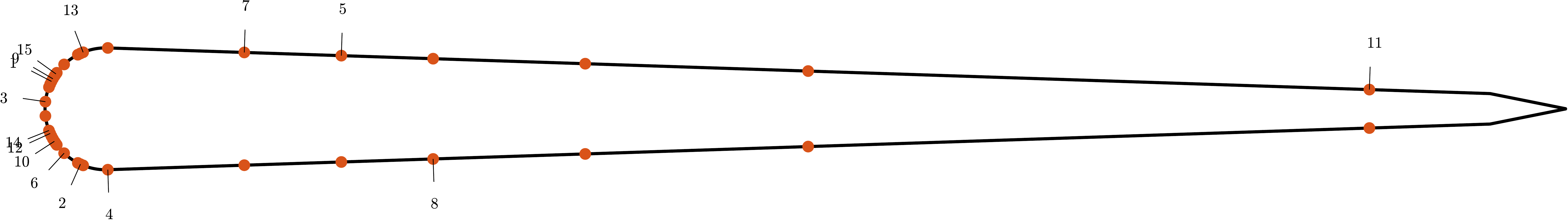}
		\subcaption{}
		\label{fig:locShearSensors}
	\end{subfigure}
	\caption{Optimal sensor locations for the shear stress measurements for the ``\EXPa'' in \cref{fig:numFishShearSensors} experiment and ``\EXPb'' experiment in \cref{fig:locShearSensors}.}
	\label{fig:sequentialPlacementShear}
\end{figure}

We perform sequential placement of $20$ sensors. The resulting distribution of sensors is shown in~\cref{fig:sequentialPlacementShear}. In~\cref{sec:results:many:sensors} we argue that the expected utility must reach a plateau when placing many sensors using the Bayesian central limit theorem. For shear stress sensors we observe that the convergence is slower compared to the pressure gradient sensors. We conclude that the information gain per shear stress sensor placed is lower as for the pressure gradient sensors.

The posterior density obtained for both experiments is less informative when using the same number of sensors. This indicates that shear is a less informative quantity yielding a slower convergence of the posterior.

\section{Discussion}
\label{sec:conclusion}

We present a study of the optimal sensor locations on a self-propelled swimmer for detecting the number and location of a leading group of  swimmers.
This optimization combines Bayesian experimental design with large scale simulations of the two dimensional Navier-Stokes equations. Mimicking the function of sensory organs  in real fish, we used the shear stress and pressure gradient on the surface of the fish to determine the sensor feedback generated by a disturbance in the flow field.

The optimization was performed for different configurations of fish, ranging from a simple leader-follower configuration with two fish, to a group of up to eight fish leading a single follower. We regarded two types of information, the number of fish in the leading group and the relative location of the leading group. We find that although the general shape of the utility function varies between the two objectives, the preferred location of the first sensor on the head of the fish is consistent. Furthermore, we find that the objective is only weakly influenced when varying the number of members in the leading group.

We perform a sequential sensor placement and find that the utility converges to a constant value and thus we can conclude that few sensors suffice to infer the quantities of the surrounding flow. 
Indeed we find that  the optimal sensor locations correspond to a posterior distribution that is strongly peaked around the true value of the quantity of interest. 

In summary, we find  that changing the number of fish in the leading group does not influence the followers ability to infer the location of the leading group, for groups with small size. Furthermore we were able to show that choosing the locations for the measurements in a systematic way we are able to infer the number of fish in the leading group and the location of our agent to high accuracy.
Ongoing work examines the inclusion of further parameters such as the motion of the swimmers in optimally detecting disturbances from neighboring fish.

\vspace{6pt} 



\authorcontributions{Conceptualization, Georgios Arampatzis and Petros Koumoutsakos; Data curation, Pascal Weber; Formal analysis, Pascal Weber and Georgios Arampatzis; Funding acquisition, Petros Koumoutsakos; Investigation, Pascal Weber; Methodology, Georgios Arampatzis, Siddhartha Verma and Costas Papadimitriou; Project administration, Petros Koumoutsakos; Resources, Petros Koumoutsakos; Software, Pascal Weber and Guido Novati; Supervision, Georgios Arampatzis and Petros Koumoutsakos; Validation, Pascal Weber and Guido Novati; Visualization, Pascal Weber and Georgios Arampatzis; Writing – original draft, Pascal Weber; Writing – review \& editing, Pascal Weber, Georgios Arampatzis, Guido Novati, Siddhartha Verma, Costas Papadimitriou and Petros Koumoutsakos.}

\funding{We would like to acknowledge the computational time at Swiss National Supercomputing Center (CSCS) under the project s929. We gratefully acknowledge support from the European Research Council (ERC) Advanced Investigator Award (No. 341117).}


\conflictsofinterest{The authors declare no conflict of interest.} 



\appendixtitles{no} 
\appendix

\section{Configurations}\label{app:configurations}

The configuration used for the number of fish experiment. For the configurations with three rows the vertical extent $y\in [0,0.5]$ was discretized using $2048$ gridpoints, for the ones with four rows it was extended to $y\in [0,0.75]$ and discretized using $3072$ gridpoints.

\begin{figure}[H]
	\centering
	\includegraphics[width=\textwidth]{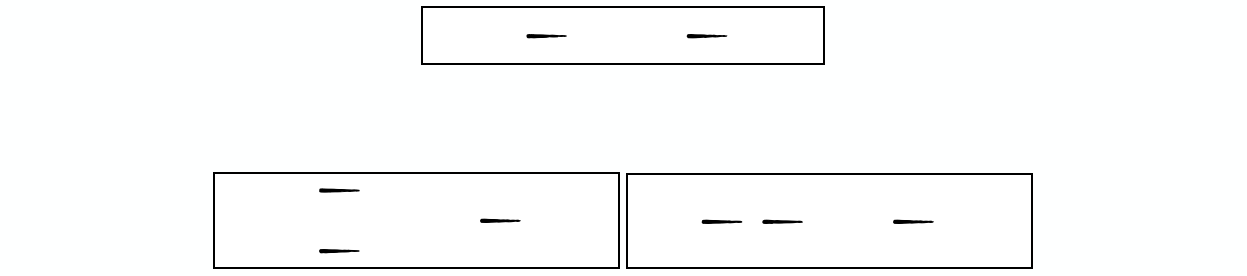}
	\label{fig:conf2fish}
	\caption{Configurations for two fish.}
\end{figure}

\begin{figure}[H]
	\centering
	\includegraphics[width=\textwidth]{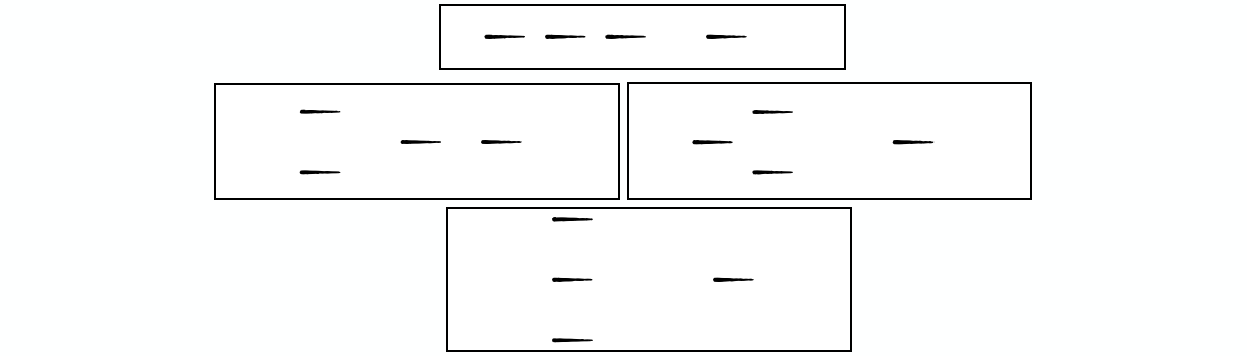}
	\label{fig:conf3fish}
	\caption{Configurations for three fish.}
\end{figure}

\begin{figure}[H]
	\centering
	\includegraphics[width=\textwidth]{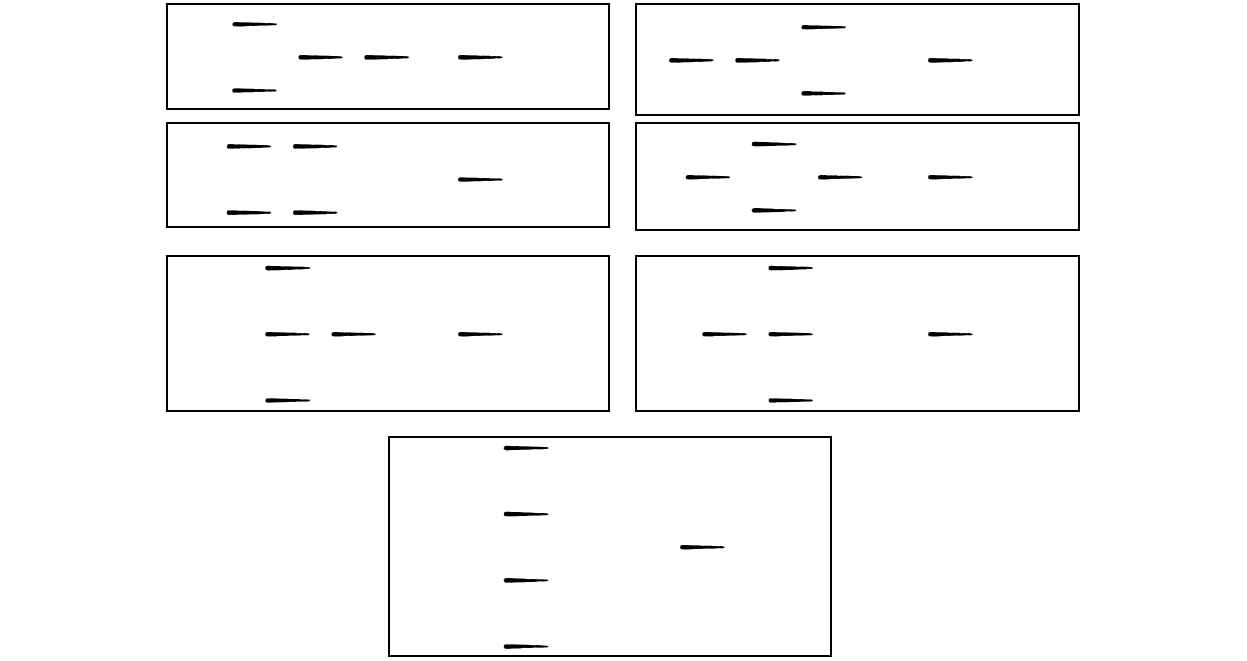}
	\label{fig:conf4fish}
	\caption{Configurations for four fish.}
\end{figure}

\begin{figure}[H]
	\centering
	\includegraphics[width=\textwidth]{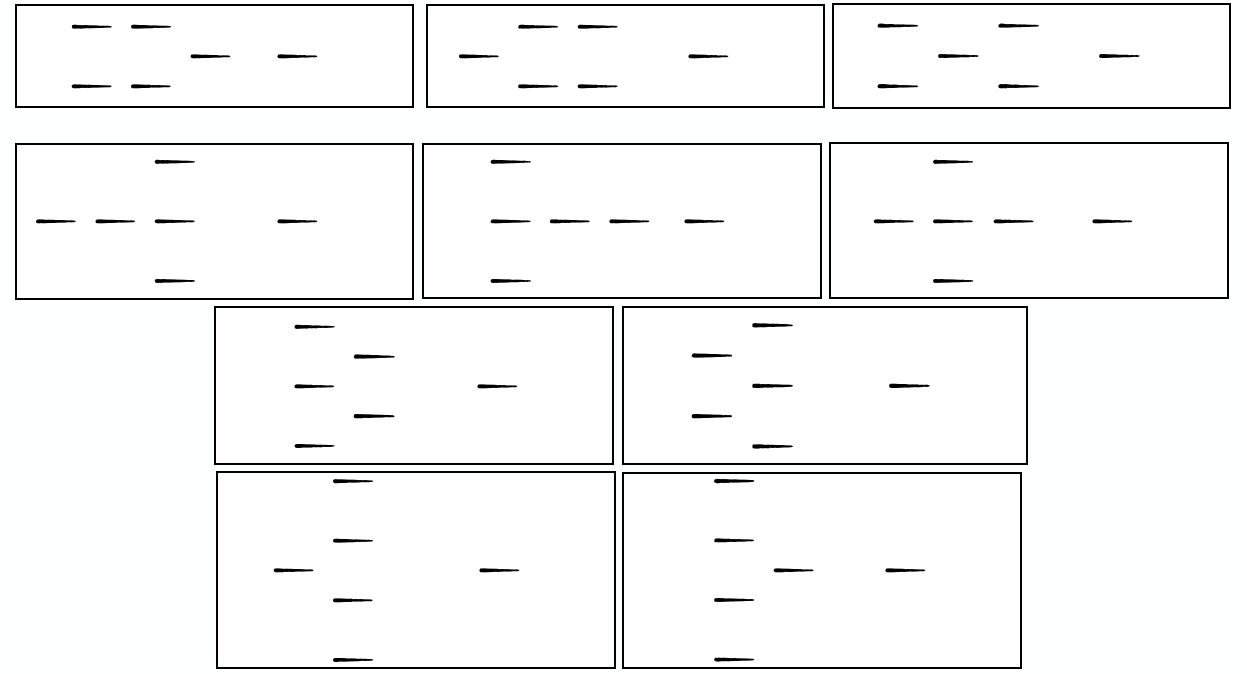}
	\label{fig:conf5fish}
	\caption{Configurations for five fish.}
\end{figure}

\begin{figure}[H]
	\centering
	\includegraphics[width=\textwidth]{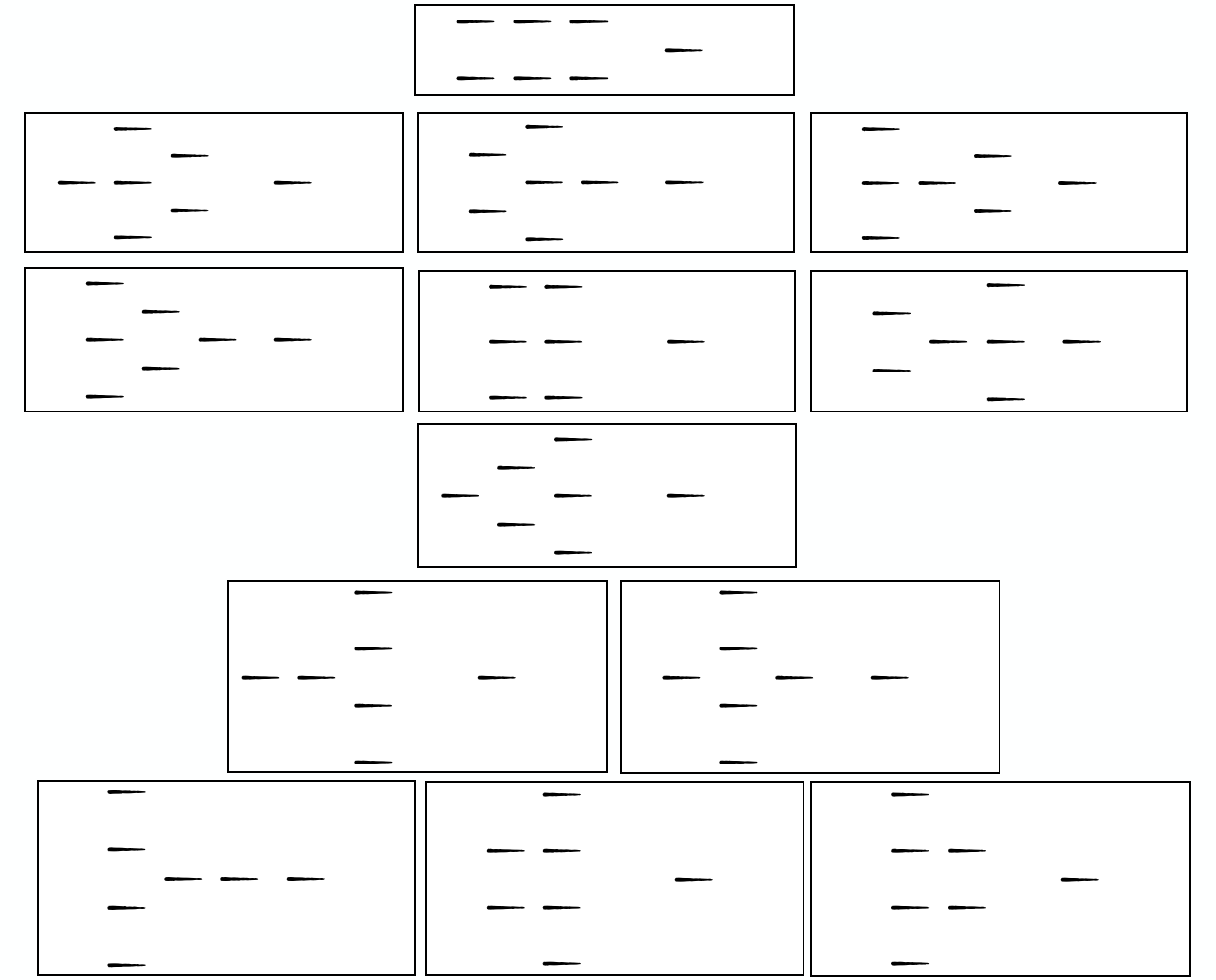}
	\label{fig:conf6fish}
	\caption{Configurations for six fish.}
\end{figure}

\begin{figure}[H]
	\centering
	\includegraphics[width=\textwidth]{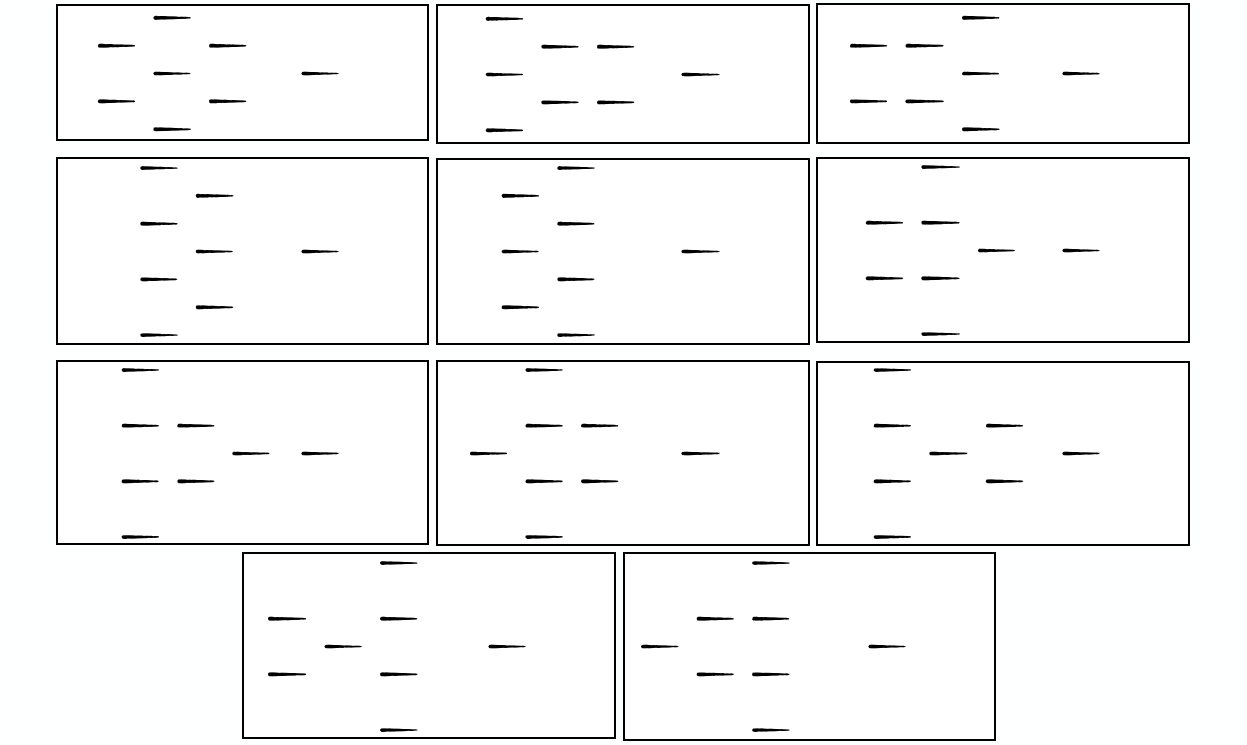}
	\label{fig:conf7fish}
	\caption{Configurations for seven fish.}
\end{figure}

\begin{figure}[H]
	\centering
	\includegraphics[width=\textwidth]{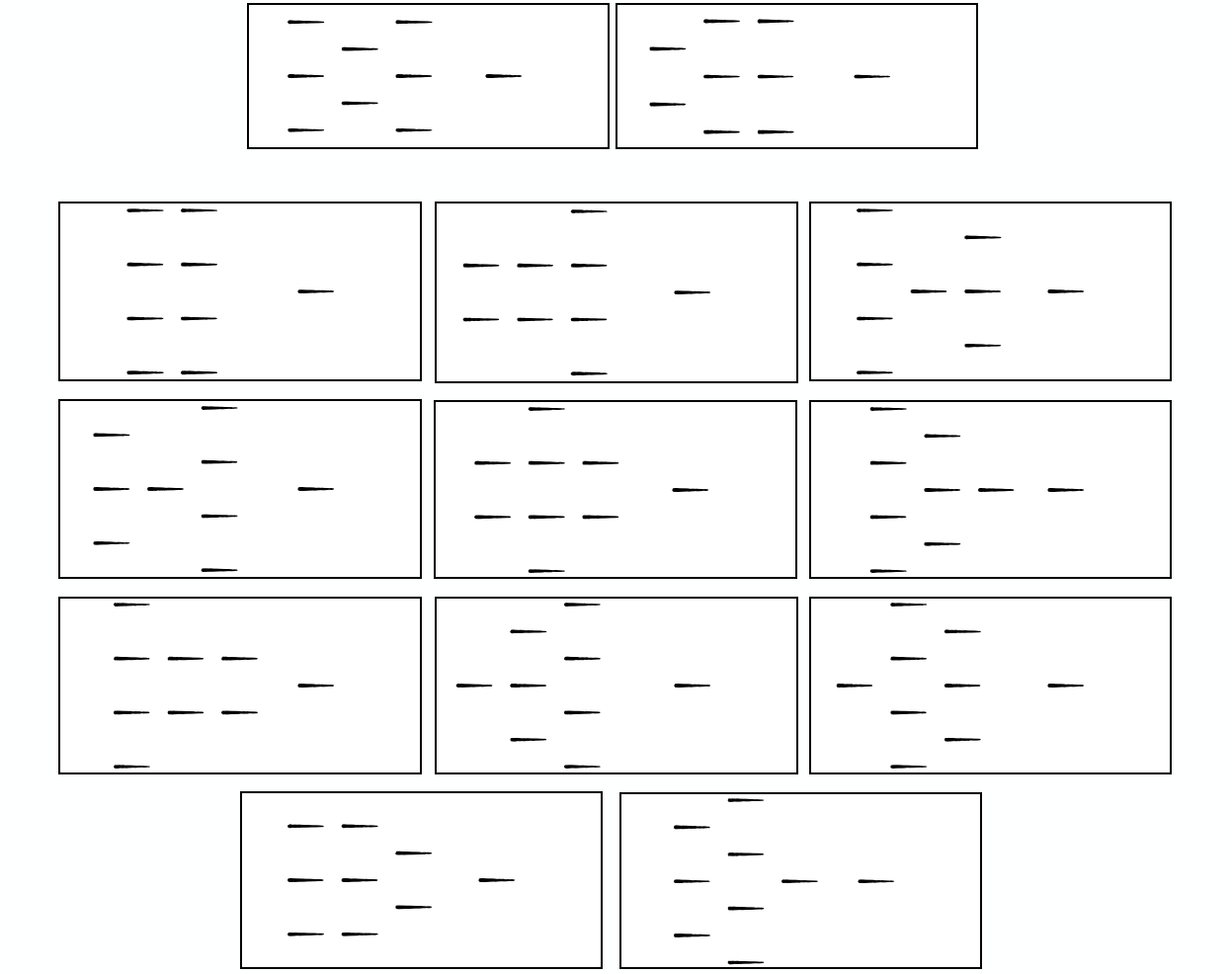}
	\label{fig:conf8fish}
	\caption{Configurations for eight fish.}
\end{figure}

\section{Posterior is not symmetric} \label{app:posterior:symmetry}

The posterior in~\cref{fig:NumFishPosterior} has no mirror symmetry along the diagonal axis. This observation indicates that the posterior is not symmetric with respect to an exchange of the parameter $\PRM$ we try to infer and the parameter used in the simulation $\PRM_{\text{true}}$. In the following we want to formally examine this observation. For sake of brevity we neglect the influence on the sensor location $\SNS$. The posterior for some $\PRM_i$ with $i=1,\dots,8$ is given by Bayes rule

\begin{equation}
p(\PRM_i|\MS)\propto p(\PRM_i)p(\MS|\PRM_i)\PERIOD
\end{equation}
In \cref{sec:utilityDiscrete} we showed that for multiple configurations $\CFG_{i,\ell}$ for a given number of fish $\PRM_i$ the likelihood is an equally weighted mixture of Gaussian distributions. Further assuming an uniform prior distribution, we have

\begin{equation}
p(\PRM_i)p(\MS|\PRM_i)=\frac{1}{8}\frac{1}{n_{i}}\sum\limits_{\ell=1}^{n_i}\frac{1}{\sqrt{2\pi\sigma^2}}\exp\left(-\frac{(\MS-\mathbf{F}(\CFG_{i,\ell}))^2}{2\sigma^2}\right)\PERIOD
\end{equation}
Let us take $\MS=\mathbf{F}(\CFG^*)$ for some fixed configuration $\CFG^*$ belonging to some number of fish $\PRM^*$. If we replace $\CFG^*\to \CFG_{i,\ell^*}$ for some fixed configuration $\CFG_{i,\ell^*}$ corresponding to $\PRM_i$ and $\CFG_{i,\ell}\to \CFG_{*,\ell}$ for the number of fish $\PRM^*$, we realize that for $n_i,n_*\ne 1$ the exponents in the resulting sum of Gaussian densities are different

\begin{equation}
\frac{1}{8}\frac{1}{n_{i}}\sum\limits_{\ell=1}^{n_i}\frac{1}{\sqrt{2\pi\sigma^2}}\exp\left(-\frac{(\mathbf{F}(\CFG^*)-\mathbf{F}(\CFG_{i,\ell}))^2}{2\sigma^2}\right)\ne \frac{1}{8}\frac{1}{n_{*}}\sum\limits_{\ell'=1}^{n_*}\frac{1}{\sqrt{2\pi\sigma^2}}\exp\left(-\frac{(\mathbf{F}(\CFG_{i,\ell^*})-\mathbf{F}(\CFG_{*,\ell'}))^2}{2\sigma^2}\right)\COMMA
\end{equation} 
We remark that in case we have only one configuration $n_i=1$ for all $i=1,\dots,8$ the contrary is true and the posterior becomes symmetric with respect to an exchange of the parameter $\PRM$ we try to infer and the parameter used in the simulation $\PRM_{\text{true}}$.




\externalbibliography{yes}
\bibliography{bibliography.bib}



\end{document}